\documentclass[prd,aps,twocolumn,nofootinbib,superscriptaddress,eqsecnum,floatfix,preprintnumbers,amsmath,amssymb,
nofootinbib,longbibliography]{revtex4-1}

\usepackage{amssymb,amsmath}
\usepackage{epsfig}
\usepackage[dvipsnames]{xcolor}
\usepackage[utf8]{inputenc}
\usepackage{stmaryrd}
\usepackage{mathrsfs}
\usepackage{mathalfa}
\usepackage{accents}
\usepackage[normalem]{ulem}
\usepackage{enumitem}

\usepackage{graphicx}

\DeclareMathOperator\erf{erf}
\DeclareMathOperator\erfc{erfc}

\DeclareMathOperator\sgn{sgn}

\newcommand{\R}{\mathcal{R}}

\newcommand{\pa}{\partial}

\newcommand{\be}{\begin{equation}}
\newcommand{\ee}{\end{equation}}
\newcommand{\BS}{\begin{split}}
\newcommand{\ES}{\end{split}}
\newcommand{\bea}{\begin{eqnarray}}
\newcommand{\eea}{\end{eqnarray}}
\newcommand{\ba}{\begin{equation}\begin{aligned}}
\newcommand{\ea}{\end{aligned}\end{equation}}

\newcommand{\beg}{\begin{gather*}}
\newcommand{\eng}{\end{gather*}}
\newcommand{\hh}{,\hspace{0.5cm}}
\newcommand{\hhh}{,\hspace{0.2cm}}

\newcommand{\lap}{\triangle}

\newcommand{\n}[1]{\label{#1}}

\newcommand{\ts}[1]{{\boldsymbol{#1}}}

\def\XXint#1#2#3{{\setbox0=\hbox{$#1{#2#3}{\int}$ }
\vcenter{\hbox{$#2#3$ }}\kern-.6\wd0}}

\usepackage[makeroom]{cancel}
\usepackage[caption=false]{subfig}
\usepackage[colorlinks=true,
            citecolor=green,
            linkcolor=red,
            filecolor=cyan,
            urlcolor=magenta,
            backref=false]{hyperref}
\newcommand{\hb}{\hat{b}}

\newcommand{\zz}{\mathcal{Z}}

\newcommand{\ox}[1]{\!\stackrel{o}{#1}}

\begin{document}

\title{Nonlocal modification of the Kerr metric}

\author{Valeri P. Frolov}%
\email[]{vfrolov@ualberta.ca}
\affiliation{Theoretical Physics Institute, Department of Physics,
University of Alberta,\\
Edmonton, Alberta, T6G 2E1, Canada
}
\author{Jose Pinedo Soto}
\email[]{pinedoso@ualberta.ca}
\affiliation{Theoretical Physics Institute, Department of Physics,
University of Alberta,\\
Edmonton, Alberta, T6G 2E1, Canada
}

\begin{abstract}
In the present paper, we discuss a nonlocal modification of the Kerr metric. Our starting point is the Kerr-Schild form of the Kerr metric $g_{\mu\nu}=\eta_{\mu\nu}+\Phi l_{\mu}l_{\mu}$. Using Newman's approach we identify a shear free null congruence $\ts{l}$ with the generators of the null cone with apex at a point $p$ in the complex space. The Kerr metric is obtained if the potential $\Phi$ is chosen to be a solution of the flat Laplace equation
for a point source at the apex $p$. To construct the nonlocal modification of the Kerr metric we modify the Laplace operator $\lap$ by its nonlocal version $\exp(-\ell^2\lap)\lap$. We found the potential $\Phi$ in such an infinite derivative (nonlocal) model and used it to construct the
sought-for nonlocal modification of the Kerr metric. The properties of the rotating black holes in this model are discussed. In particular, we derived and numerically solved the equation for a shift of the position of the event horizon due to nonlocality.

\hfill {\scriptsize Alberta Thy 5-23}
\end{abstract}

%\today

\maketitle

\section{Introduction}

The Kerr metric discovered by Roy Kerr \cite{Kerr} is the most general vacuum solution of the Einstein equations describing a stationary rotating black hole in an asymptotically flat spacetime. It is widely used in astrophysics both for the description of the gravitational field of stellar mass and supermassive black holes as well as in the study of the coalescence of black holes. The properties of the Kerr metric are well known and are described in a number of books (see e.g. \cite{Carter:1973rla,MTW,chandrasekhar,Frolov:1998wf,kramer,Neill} and references therein). The Kerr metric, besides two commuting Killing vectors generating time translation and rotation, possesses a hidden symmetry. Namely, it has a so called closed conformal Killing-Yano tensor which generates a second rank Killing tensor \cite{Penrose,Floyd}. As a result, the geodesic equations of motion of a particle in the Kerr spacetime are completely integrable and the additional quadratic in momentum integral of motion (Carter's constant \cite{PhysRev.174.1559}) is constructed by using the Killing tensor. (A comprehensive discussion of the hidden symmetries in black hole spacetimes and further references can be found in \cite{Living_Frolov:2017kze}.)

Another remarkable property of the Kerr metric (as well as of its charged version, the Kerr-Newman metric \cite{Kerr-Newman,Kerr-Newman_1}) is that it can be written in the Kerr-Schild  form \cite{Kerr_Schild}
\begin{equation}\n{KS_FORM}
g_{\mu\nu}=\eta_{\mu\nu}+\Phi l_{\mu}l_{\nu}\, ,
\end{equation}
where $\eta_{\mu\nu}$ is a flat metric, $\Phi$ is a scalar field, and $\ts{l}$ is a  tangent vector to a shear-free geodesic null congruence.
It has been shown that these solutions of the Einstein equations can be obtained by complex coordinate transformations from the Schwarzschild metric \cite{Newman:1965tw,Newman:1973afx}. In particular, the potential $\Phi$ for the Kerr metric can be obtained as a solution of the Laplace equation in flat coordinates $(X,Y,Z)$
\begin{equation}\n{lap}
\lap\Phi=4\pi j\, ,
\end{equation}
with a point-like source $j$ located at the complex coordinate $Z+ia$, where $a$ is the rotation parameter of the Kerr black hole \cite{Israel,Kaiser_2003}. A comprehensive review of the Kerr-Schild metrics and complex space approaches can be found in \cite{Adamo:2014baa}.

More recently, the Kerr-Newman representation of the spacetime geometry received further development and modifications in the so-called double copy formalism. The main idea of this approach is based on the observation that for the metrics which allow the Kerr-Schild representation the non-linear Einstein equations can be reduced to the linear equations for Maxwell and scalar fields. This observation can be used to simplify calculations of gravity scattering amplitudes by reducing this problem to the calculation of the Yang–Mills amplitudes with a subsequent double copy prescription \cite{Bern_2010,Monteiro_2014,Luna_2015,Bah:2019sda}. At the moment there exist dozens of publications on this subject. Related references can be found e.g. in the following review articles \cite{White_2018,bern2019duality,bern2022sagex,dePaulaNetto:2023vtg}.

In this paper, we propose a model of a nonlocal modification of the Kerr metric and discuss its properties. The main idea of this approach is the following. We use the Kerr-Schild ansatz for the metric but modify the equation \eqref{lap} for the potential and write it in the form
\begin{equation}\n{GF_lap}
f(\lap)\lap\Phi=4\pi j\, ,
\end{equation}
with a specially chosen form factor function $f(z)$.
In particular, we assume that the form factor is chosen such that it does not vanish in the complex plane of $z$, and hence it has a unique inverse.
As a result, no new unphysical degrees of freedom are
present (at least at tree level). For this reason, such nonlocal (infinite derivative) theories are sometimes referred to as ``ghost-free".
Quite often the form factor satisfying these conditions is chosen in the form
\begin{equation}\n{GF_N}
f(\lap)=\exp\left[(-\ell^2 \lap)^N\right] \, .
\end{equation}
Here $N$ is a positive integer number, and $\ell$ plays the role of the fundamental length specifying a length scale at which the effects of nonlocality become important. One refers to this kind of nonlocality as to $GF_N$ model.

These kinds of models have been studied in many publications starting with the papers \cite{Tomboulis:1997gg,Moffat:2010bh,Modesto:2011kw,Biswas_2012,Modesto:2010uh,Biswas_2013} .
The main motivation for studying such models is the following. It is well known that the standard Einstein gravity theory is ultraviolet incomplete. In the classical theory, this incompleteness manifests itself in the inevitable presence of singularities both in cosmology and in the black hole interior. One can try to improve the ultraviolet behavior of the theory by adding higher orders in the derivatives of the curvature terms of the action. However, this usually results in new unphysical degrees of freedom (ghosts) arising. The interest in the infinite derivative (nonlocal) modifications of Einstein's gravity is partially motivated by the hope of overcoming this difficulty.

Solutions for the gravitational field of point-like sources in linearized ghost free gravity were obtained and studied in many papers references to which can be found e.g. in \cite{Boos_2018}.
A solution of these equations  when the source is  a rotating infinitely thin massive ring was found in \cite{Buoninfante:2018xif}.
Cosmology in the nonlocal stringy models was studied in \cite{Arefeva:2007wvo,Arefeva:2007wvo,Koshelev:2014voa}.
Exact pp-wave and gyraton type solutions in the infinite derivative gravity were discussed in \cite{Kilicarslan:2019njc,Dengiz:2020xbu,Kolar:2021uiu}.
Additional references can be found in the reviews \cite{Boos:2020qgg,Modesto:2017sdr,Buoninfante:2019zws,Heredia:2021pxp,Kolar:2022kgx,Buoninfante:2022ild}.

In this paper, we consider the following modification of the Kerr solution, which for briefness we call the "nonlocal Kerr metric". We start with the Kerr-Schild form \eqref{KS_FORM} of the metric. We keep the same shear-free, geodesic null congruence $\ts{l}$ and the same point like source $j$ in the complex space as for the Kerr solution. However, we modify the potential $\Phi$ and choose it to be a solution of the equation \eqref{GF_lap} with a specially chosen (ghost free) form factor. Our goal is to obtain such a nonlocal Kerr metric and to study its properties.

Let us stress that such a metric certainly is not a solution to the exact infinite derivative equations, which are highly nonlinear \cite{Kolar:2021rfl}. At the same time the obtained nonlocal Kerr metric, written in coordinates similar to the Boyer-Lindquist coordinates, is non-linear in the mass parameter. It describes a stationary axisymmetric black hole which in several aspects differs from the Kerr spacetime. Written in the Kerr-Schild form \eqref{KS_FORM} this metric, similarly to the Kerr solution, looks like a linear perturbation of the flat spacetime.
However, the coordinate transformation, required to present the metric in Boyer-Lindquist form non-linearly depends on the scalar function $\Phi$. For this reason, even for the weak nonlocality, the nonlocal Kerr metric cannot be obtained by a small change of the mass parameter $M$ in the Kerr metric, for example by taking its slightly dependent on the radial and angle coordinates.

The paper is organized as follows. In section~\ref{S2} we discuss the Kerr-Schild form of the metric and describe different coordinates which are used later in the paper. Section~\ref{S3} discusses a definition of the delta function in the complex space and contains the derivation of the potential $\Phi$, which is a solution of the Poisson equation with a complex delta function. A similar solution for an infinite derivative modification of the Poisson equation with the same point-like source in the complex space is derived in section~\ref{S4}. This section also contains a discussion of the properties of the nonlocal potential. In section~\ref{S5} we use the obtained nonlocal potential to recover the nonlocal modification of the Kerr metric. The spacetime structure of such a black hole, including the shift of the event horizon due to nonlocality, is also discussed in section~\ref{S5}. In section~\ref{S6} we discuss a limiting case of a nonrotating nonlocal black hole.
Section~\ref{S7} contains a discussion of the obtained results. Technical details and calculations required for the derivation of the equation for the event horizon shift are discussed in the appendix.

\section{Kerr metric and its Kerr-Schild form}
\n{S2}
\subsection{Kerr metric}

The Kerr metric describing a vacuum stationary rotating black hole written in the Boyer-Lindquist coordinates is
\begin{equation}\label{Kerr}
\begin{split}
dS^2 =& -\left( 1-\frac{2Mr}{\Sigma}\right) dt^2
-\frac{4Mar\sin^2\theta}{\Sigma} dt d\phi\\
+&\left(r^2+a^2+\frac{2M a^2 r}{\Sigma} \sin^2\theta\right)\sin^2\theta d\phi^2\\
+&\frac{\Sigma}{\Delta} dr^2 +\Sigma d\theta^2\, ,\\
&\Sigma=r^2+a^2\cos^2\theta\hh \Delta=r^2-2Mr+a^2 \,  .
\end{split}
\end{equation}
Here $M$ is the black hole mass, and $a$ is its rotation parameter. This metric has two commuting Killing vectors $\ts{\xi}_{(t)}=\pa_t$ and $\ts{\xi}_{(\phi)}=\pa_{\phi}$\footnote{Many useful relations for the Kerr metric and its Kerr-Schild form can be found in \cite{Visser:2007fj}.}.

The projection of the metric \eqref{Kerr} along the orbits of the Killing vectors determines a smooth two-dimensional space $S$ with metric \cite{Geroch_2}
\begin{equation}
dl^2=\frac{\Sigma}{\Delta} dr^2 +\Sigma d\theta^2\, .
\end{equation}
The Killing vectors $\ts{\xi}_{(t)}$ and $\ts{\xi}_{(\phi)}$
satisfy the following circularity condition (see e.g. \cite{Carter:1973rla,kramer,Frolov:1998wf})
\begin{equation}\label{CIRC}
\xi_{(\phi)\,[\alpha}\xi_{(t)\beta}\xi_{(t)\gamma ;\delta]}=
\xi_{(t) [\alpha}\xi_{(\phi)\beta}\xi_{(\phi)\gamma ;\delta]}=0\, .
\end{equation}
These relations are necessary and sufficient conditions for the 2-flats orthogonal to $\ts{\xi}_{(t)}$ and $\ts{\xi}_{(\phi)}$ to be integrable. Let us denote by $\Gamma$ the two-dimensional span of the Killing vectors  $\ts{\xi}_{(t)}$ and $\ts{\xi}_{(\phi)}$.  Then, the circularity condition implies that $\Gamma$ is orthogonal to $S$.

\subsection{Coordinates}

In what follows we shall use several different coordinate systems. Let us describe  them in this section.

Let us first note that for $M=0$ the Riemann curvature of the Kerr metric vanishes and the metric (\ref{Kerr}) takes the form
\begin{equation}\label{Flat}
\begin{split}
d\ox{s}{\!}^2&= - d{t}^2+dh^2\, ,\\
dh^2&=
\frac{\Sigma}{r^2+a^2} dr^2 +\Sigma d\theta^2+(r^2+a^2) \sin^2\theta d\phi^2 \, .
\end{split}
\end{equation}
In this limit the metric \eqref{Flat} is nothing but the Minkowski metric and
its spatial part $dh^2$ is flat as well. We denote by $(X,Y,Z)$ standard Cartesian coordinates in this 3D space.
Then it is easy to check the coordinates $(r,\theta,\phi)$ are related to these Cartesian coordinates as follows
\begin{equation}\n{OBLATE}
  \begin{split}
    X &= \sqrt{r^2 + a^2}\sin\theta \cos\phi \, , \\
    Y &= \sqrt{r^2 + a^2}\sin \theta \sin \phi  \, ,\\
    Z &= r \cos\theta \, .
  \end{split}
\end{equation}
The coordinates $(r,\theta,\phi)$ are nothing but standard oblate spheroidal coordinates taking the following values $r\ge 0$, $\theta\in [0,\pi]$, $\phi\in [0,2\pi]$.
For $r>0$ the surfaces $r=$const are oblate ellipsoids.
Figure~\ref{F1} shows the coordinate lines of the oblate spheroidal coordinates $(r,\theta)$ in the plane $Y=0$ ($\phi=0$).

\begin{figure}[!hbt]
    \centering
      \includegraphics[width=0.45\textwidth]{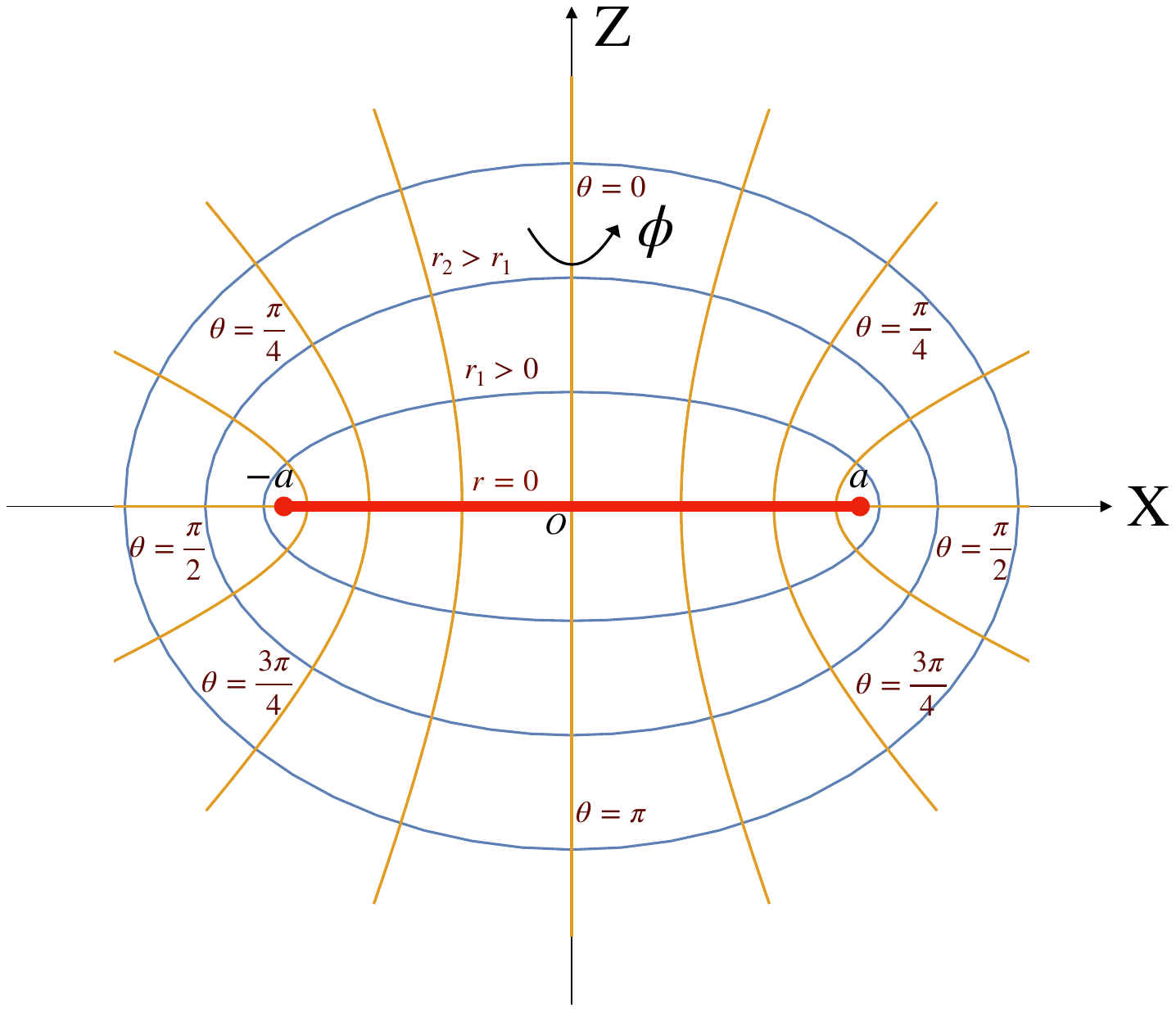}
    \caption{\n{F1} Coordinate lines of the oblate spheroidal coordinates $(r,\theta)$ in the plane $Y=0$ ($\phi=0$)}
\end{figure}

For $r=0$ and $\theta\in [0,\pi]$, $\phi\in [0,2\pi]$ one has a disc $\mathcal{D}$ of radius $a$ located in the $Z=0$ plane. The coordinate $\theta$ is discontinuous on the disc. For $(0,\pi/2)$ the coordinate $\theta$ covers the upper part of the disc, while for  $(\pi/2,\pi)$, it covers the lower part of it.
The boundary $\pa\mathcal{D}$ of this disc  is a ring of radius $a$.
Equations $\theta=0$ and $\theta=\pi$ describe the axis of symmetry $X=Y=0$. For $\theta=0$, $Z=r$  is positive, while for $\theta=\pi$, $Z=-r$  is negative.

The third type of coordinates in the flat 3D space which will be also used in the paper are the cylindrical coordinates $(\rho,z,\phi)$ related to Cartesian coordinates $(X,Y,Z)$ as
\begin{equation}
\rho=\sqrt{X^2+Y^2}\hh z=Z\, .
\end{equation}
In these coordinates the flat 3D metric is
\begin{equation}
 dh^2 = d\rho^2 + \rho^2 d\phi^2 + dz^2 \, .
\end{equation}
The cylindrical coordinates are related to the oblate spheroidal coordinates as follows
\begin{equation}\label{cyltosphero}
\begin{split}
    \rho &= \sqrt{r^2 +a^2}\sin\theta\hh z = r\cos\theta \, .
\end{split}
\end{equation}
The equation of the ring in cylindrical coordinates is $\rho=a$, $z=0$.

\begin{figure}[!hbt]
    \centering
      \includegraphics[width=0.4\textwidth]{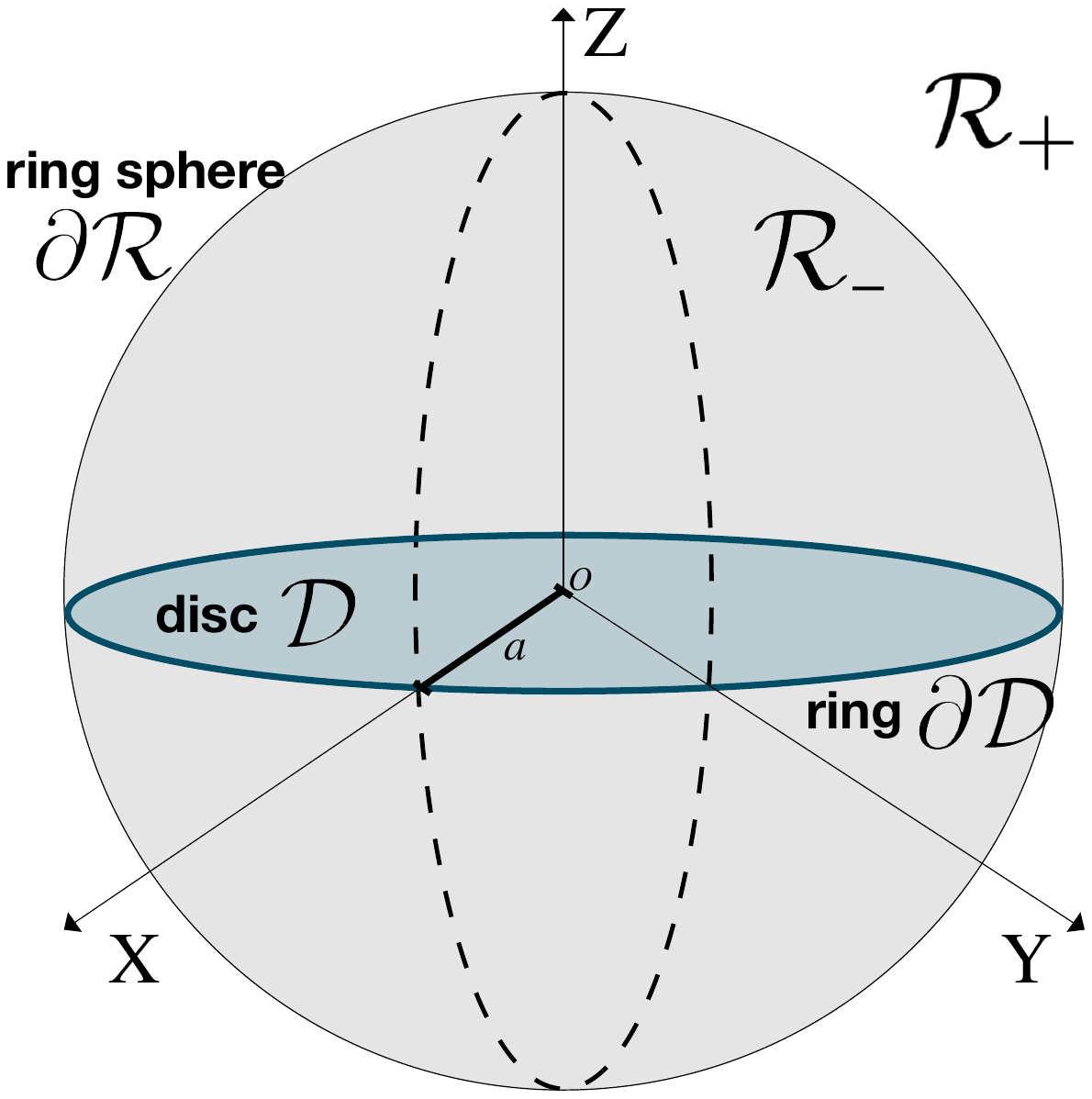}
    \caption{\n{F2} The ring $\pa\mathcal{D}$, the disc $\mathcal{D}$ and the ring sphere $\pa\mathcal{R}$.}
\end{figure}

Finally, let us introduce the forth type of the coordinates. For this purpose we define a new coordinate, $y$, related to the angle $\theta$ as follows
\begin{equation}
y=a\cos\theta.
\end{equation}
The equation of the disc $\mathcal{D}$ in $(r,y,\phi)$ coordinates is $r=0$, $y\in(-a,a)$, and $\phi\in(0,2\pi)$. The equations $r=0$, $y=0$ describe its boundary, the ring $\pa\mathcal{D}$, see Figure~\ref{F2}. This figure also shows a sphere $\pa\mathcal{R}$ of radius $a$. On its surface $r=|y|$ and $y\in (-a,a)$.
Inside the sphere  $\pa\mathcal{R}$ (in the region $\R_-$) one has $r<|y|$, while outside (in the region $\R_+$) one has $r>|y|$.

The flat metric $dh^2$ in the coordinates $(r,y,\phi)$ is
\begin{equation}\n{RY}
\begin{split}
dh^2&=\Sigma\left( \dfrac{dr^2}{\Delta^0_r}+\dfrac{dy^2}{\Delta^0_y}\right)+\dfrac{\Delta^0_r\Delta^0_y}{a^2}d\phi^2\, ,\\
\Sigma&=r^2+y^2\hhh \Delta^0_r=r^2+a^2\hhh \Delta^0_y=a^2-y^2\, .
\end{split}
\end{equation}
One can see that the metric coefficients in \eqref{RY} are simple rational functions of $r$ and $y$
and the coordinates $r$ and $y$ enter this metric in a quite symmetric way\footnote{These coordinates and their generalizations in the higher  dimensions are naturally connected with the hidden symmetries of the black hole metrics. In fact they  are eigenvalues of the rank two Killing tensor generating the hidden symmetry. For more details see e.g. \cite{FrolovZelnikov:2011, Living_Frolov:2017kze}. }.

\subsection{Kerr-Schild form}

Let us consider the following 1-form
\begin{equation}
l_{\mu}dx^{\mu}=-dt +\epsilon \frac{\Sigma}{\Delta^0_r}dr-\dfrac{\Delta^0_y}{a}d\phi\, ,
\end{equation}
where $\epsilon=\pm1$.
We define a metric
\begin{equation}\n{KSP}
d{s}^2=d\ox{s}{\!}^2+\Phi(l_{\mu}dx^{\mu})^2\, ,
\end{equation}
where $\Phi=\Phi(r,\theta)$ is some function. Then the following statements are valid for each of the metrics $d{s}^2$ and $d\ox{s}{\!}^2$. In other words, these statements are valid for an arbitrary function $\Phi$, including $\Phi=0$:
\begin{itemize}
\item The contravariant components of the vector $\ts{l}$ in $(t,r,\theta,\phi)$ coordinates  are $l^{\mu}=\left(1,\epsilon,0,-\dfrac{a}{r^2+a^2}\right)$;
\item $\ts{l}$ is a null vector $\ts{l}^2=l_{\mu}l^{\mu}=0$;
\item Vectors $\ts{l}$ are tangent vectors to incoming (for $\epsilon=-1$) or outgoing (for $\epsilon=-1$) null geodesics in the affine parameterization,   $l^{\nu}l^{\mu}_{\ ;\nu}=0$.
\item $l^{\mu}_{\ ;\mu}=\epsilon\dfrac{2r}{\Sigma}$;
\item $l_{(\mu ;\nu)}l^{(\mu ;\nu)}-\frac{1}{2}(l^{\mu}_{\ ;\mu})^2=0$\, .
\end{itemize}

The last property implies that the congruence of null vectors $\ts{l}$ is shear-free (for more details see e.g. \cite{sommers1976,Frolov1979}).
Such a null geodesic congruence is related to the light cones with apex on the world-line in the complex space. The twist is a measure of how far the complex world-line is from the real slice \cite{Newman_2004}.

Let us denote
\begin{equation} \n{VVV}
V=(\ts{\xi}_{(t)}\cdot\ts{\xi}_{(\phi)})^2-\ts{\xi}_{(t)}^2\ts{\xi}_{(\phi)}^2\, .
\end{equation}
For the metric \eqref{KSP} this quantity is
\begin{equation} \n{VP}
V=\dfrac{\Delta^0_y}{a^2}(\Delta^0_r-\Sigma\Phi)\, .
\end{equation}

It is easy to check that for a special choice of the function $\Phi$
\begin{equation} \n{Phi0}
\Phi_0=\frac{2Mr}{\Sigma} \, ,
\end{equation}
the metric $d{s}^2$ given by \eqref{KSP} is Ricci flat, and in fact, it coincides with the Kerr metric. In order to prove this it is sufficient to make the following coordinate transformation
\begin{equation}\label{BL}
\begin{split}
t =& t_{BL} - \epsilon \int\frac{2Mr}{\Delta}dr\, ,\\
\phi=& - \phi_{BL} + \epsilon \int\frac{2Mar}{(r^2+a^2)\Delta}dr\, ,
\end{split}
\end{equation}
where $\Delta$ is defined in (\ref{Kerr}). These coordinates are chosen so that the non-diagonal components
$g_{rt_{BL}}$  and $g_{r\phi_{BL}}$ of the metric $d{s}^2$ vanish. One can check that the metric $d{s}^2$ written in the $(t_{BL},r,\theta,\phi_{BL})$ coincides with the Kerr metric $dS^2$, provided one identifies the coordinates $t_{BL}$ and $\phi_{BL}$ in $d{s}^2$ with the standard Boyer-Lindquist coordinates $t$ and $\phi$ in the metric (\ref{Kerr}) \footnote{Let us emphasize that there exists quite important difference between $(t,\phi)$ and $(t_B,\phi_B)$ coordinates. Namely, the Boyer-Lindquist coordinates cover only the exterior of the black hole, that is the domain outside the event horizon, while coordinates $(t,\phi)$ can "penetrate" into the interior of the  black and white holes.}.

Carter \cite{Carter:1973rla} showed that if the circularity conditions \eqref{CIRC} are satisfied, the event horizon of an arbitrary stationary axially-symmetric black hole coincides with the Killing horizon. The latter is the set of points where
\begin{equation}
V=0\, .
\end{equation}
For the Kerr metric this condition implies that
\begin{equation}
r=r_H=M+\sqrt{M^2-a^2}\, .
\end{equation}
This relation determines the position of the event horizon of the Kerr black hole.

\section{Potential \texorpdfstring{$\Phi_0$}{Phi0} and a point charge in complex space}
\n{S3}

\subsection{Complex delta function}

Let us consider the scalar function $\Phi_0$ given by (\ref{Phi0}) in flat spacetime with the metric (\ref{Flat}). It is easy to check that it satisfies the Laplace equation
\begin{equation}\n{LP}
\lap\Phi_0=0\, ,
\end{equation}
where $\lap$ is the standard 3D flat Laplace operator which
takes the following form in Cartesian coordinates
\begin{equation}
\lap=\pa_X^2+\pa_Y^2+\pa_Z^2\, .
\end{equation}
In fact, $\Phi_0$ is a very special solution of (\ref{LP}) which has a point-like source in the complex space. Namely, it can be written in the following form
\begin{equation}
\Phi_0=-8\pi M \Re(G_0(X,Y,Z+ia))\, ,
\end{equation}
where $G_0(X,Y,Z+ia)$ is an analytical extension in the complex domain of the fundamental solution of the Laplace equation \cite{Kaiser_2003}.

To obtain the solution $G_0(X,Y,Z+ia)$ let us, following \cite{Kaiser_2003,Brewster_2018}, define a delta function in the complex plane. Here and later we denote
\begin{equation}\n{ZZZ}
\zz=z+ia\, ,
\end{equation}
A generalized delta function $\tilde{\delta}(\zz)$ of a complex argument $\zz$ is defined as \cite{Brewster_2018}
\begin{equation}\label{limitdelta}
    \tilde{\delta}(\zz) = \lim_{\sigma \to \infty} \frac{1}{2\pi} \int_{-\infty}^{\infty} e^{-i\zz p}e^{-p^2/2\sigma^2}dp \, .
\end{equation}
Here $\sigma$ is constant. The Gaussian exponent containing $\sigma$ is introduced to provide convergence of the integral over $p$. The prescription $\lim_{\sigma \to \infty}$ means that the limit $\sigma\to\infty$ should be taken at the end of the calculations.
It should be mentioned that this expression is divergent in the quadrants $|\Re(\zz)|\le |\Im(\zz)|$ and converges to zero everywhere else. But if both endpoints of the integration contour are in the convergent sector the definition (\ref{limitdelta}) can be used.

Let  $f (z)$ be a test function of the complex variable $z$, which is analytic throughout the complex plane and that decreases sufficiently rapidly at
large distances along the real axis. Then, as it is shown in \cite{Kaiser_2003,Brewster_2018}, the following relation is valid
\begin{equation}
    \int_{-\infty}^{\infty}  f(x) \tilde{\delta}(x-z) dx = f(z) \, .
\end{equation}
Using expression (\ref{limitdelta}) it easy to check that $\tilde{\delta}(-\zz)=\tilde{\delta}(\zz)$.

In what follows we shall be using the real part of the complex delta function
\begin{equation}\label{REdelta}
\begin{split}
  \delta_R(\zz) &=
 \frac{1}{2}( \tilde{\delta}(\zz)+\tilde{\delta}(\bar{\zz}))  \, \\
    &=\lim_{\sigma \to \infty} \frac{1}{2\pi} \int_{-\infty}^{\infty} \cos(zp)e^{ap}e^{-p^2/2\sigma^2}dp \, .
\end{split}
\end{equation}
It is easy to check that $\delta_R(z+ia)=
\delta_R(-z+ia)$. Hence this object is an even function of $z$.
Other properties of the generalized delta function and its application can be found in \cite{LINDELL,Kaiser_2003,Smagin,Berry,Brewster_2018}.

\subsection{Potential of a point source in complex space}

Using the definition of the complex delta-function one can define $G_0(X,Y,\zz)$ as a solution of the following equation
\begin{equation} \n{GF0}
\lap G_0(X,Y,\zz)=\delta(X)\delta(Y)\tilde{\delta}(\zz)\, .
\end{equation}
Here we use the notation introduced in \eqref{ZZZ}.

Denote $\vec{\rho}=(X,Y)$ and $\vec{\eta}=(\eta_X,\eta_Y)$.
Then
\begin{equation}\begin{split}\n{1_19}
\delta(X)\delta(Y)&=\frac{1}{(2\pi)^2}\int e^{-i \vec{\eta}\cdot\vec{\rho}}d^2\vec{\eta}\, ,\\
 G_0(X,Y,\zz)&=\frac{1}{(2\pi)^2}\int e^{-i \vec{\eta}\cdot\vec{\rho}} \tilde{G}_0(\vec{\eta},\zz) d^2\vec{\eta}\, .
\end{split}
\end{equation}

We use the following representation for the function $\tilde{G}_0(\vec{\eta},\zz)$
\begin{equation}\n{1_20}
\tilde{G}_0(\vec{\eta},\zz)=   \lim_{\sigma \to \infty} \frac{1}{2\pi} \int_{-\infty}^{\infty} e^{-i\zz p}e^{-p^2/2\sigma^2}\tilde{G}_0(\eta,p)\ dp\, .
\end{equation}
Then using equation (\ref{GF0}) one finds the  Fourier transform $\tilde{G}_0(\eta,p)$ of the Green function $G_0(X,Y,\zz)$
\begin{equation}\n{GGPP}
\tilde{G}_0(\eta,p)=-\frac{1}{\eta^2+p^2}\, .
\end{equation}
Here $\eta^2=\vec{\eta}^{\, 2}$

Combining these results  one gets
\begin{equation}
\begin{split}
    G_0(X,Y,\zz) = &-\frac{1}{(2\pi)^3}\int d^2\eta  e^{-i \vec{\eta}\cdot \vec{\rho}  }\, Y_0(\eta,\zz)\, ,\\
    Y_0(\eta,\zz)=&\lim_{\sigma \to \infty} \int_{-\infty}^{\infty}  dp \frac{e^{-p^2/2\sigma^2} e^{-ip \zz}}{\eta^2+p^2} \, .
\end{split}
\end{equation}
Here $\vec{\rho}=(X,Y)$.
Let $\vec{\eta}\cdot\vec{\rho} = \eta \rho \cos\phi$ and $d^2\eta = \eta d\eta d\phi$, then the integration over $\phi$ in the range $(0,2\pi)$ yields
\begin{equation}\label{BES}
   \int_0^{2\pi} d\phi  e^{-i\rho\eta\cos\phi}=2\pi J_0(\eta\rho)\, .
\end{equation}
Thus
\begin{equation}\n{315}
 G_0(\rho,\zz)=-\frac{1}{4\pi^2}\int_0^{\infty} d\eta \eta Y_0(\eta,\zz) J_0(\eta\rho)\, .
\end{equation}
This expression shows that written in the cylindrical coordinates the Green function $G_0$ does not depend on the angle $\phi$. For this reason instead of the arguments $X$ and $Y$ of the Green function we use a polar radius in the cylindrical coordinates $\rho=\sqrt{X^2+Y^2}$.

The integral over $p$ for $Y_0$ can be taken with the following result
\begin{equation}
Y_0=\frac{\pi}{\eta}\lim_{\sigma \to \infty}\exp(\frac{\eta^2}{2\sigma^2}-\eta\zz)(1-\erf{(\frac{\eta}{\sqrt{2}\sigma}))}\, .
\end{equation}
Here $\erf(z)$ is the error function of a complex variable $z$.
Its definition and properties can be found in \cite{oldham2010atlas}.

The limit $\sigma\to\infty$ can be easily taken and one gets
\begin{equation}
Y_0=\frac{\pi e^{-\eta\zz}}{\eta}\, .
\end{equation}
Using this result and expression (\ref{315}) one gets
\begin{equation}
    G_0(\rho,\zz) = - \frac{1}{4\pi} \int_0^{\infty} d\eta  e^{-\eta \zz}J_{0}(\eta \rho)\, ,
\end{equation}
which finally gives
\begin{equation}\label{verde}
   G_0(\rho,\zz)\equiv - \frac{1}{4\pi\sqrt{\rho^2 + \zz^2}} \, .
\end{equation}
It is easy to check that
\begin{equation}
\rho^2 + \zz^2=(r+ia\cos\theta)^2\, .
\end{equation}
The square root has a branch point. In what follows we use the following prescription
\begin{equation}\n{RA}
\sqrt{\rho^2 + \zz^2}=r+ia\cos\theta\hhh r\in[0,\infty]\hhh \theta\in [0,\pi]
\, .
\end{equation}
Here $(r,\theta)$ are oblate spheroidal coordinates \eqref{OBLATE}.

Hence we can write relation (\ref{verde}) as
\begin{equation}\label{G0}
   G_0(r,\theta) = - \frac{1}{4\pi}\frac{1}{r + ia\cos\theta}  = - \frac{1}{4\pi}\frac{r - ia\cos\theta}{r^2 + a^2\cos^2\theta} \, .
\end{equation}
This relation implies that
\begin{equation}\label{P0}
\Phi_0=-8\pi M\Re{\left[ G_0(r,\theta)\right]} =\frac{2Mr}{r^2 + a^2\cos^2\theta}
\end{equation}
Which correctly reproduces the expression \eqref{Phi0}.

Let us note that similar solutions  for a point source in the complex space can be  found in the Maxwell theory. Such an electromagnetic field and its properties were studied in \cite{Kaiser_2004}. Potential \eqref{P0} was also used in \cite{Eleni_2020} to construct the Newtonian analogue of the Kerr metric.

\section{Potential $\Phi$ in an infinite derivative model}
\n{S4}
\subsection{Integral representation of the nonlocal Green function}

In order to obtain the nonlocal modification of the Kerr metric we proceed as follows. At first, we calculate a nonlocal version of the potential function $\Phi_0$. To achieve this, we consider the following modification of the equation (\ref{GF0})
\begin{equation} \n{GF}
f(\lap)\lap G(X,Y,\zz)=\delta(X)\delta(Y)\tilde{\delta}(\zz)\, .
\end{equation}
Here $f$ is a form factor that is chosen so that it does not produce new (unphysical) poles. For example, one can take it in the form
\begin{equation}\n{FF}
f(\lap)=\exp[(-\ell^2\lap)^N]\hh \ell>0\, ,
\end{equation}
where $N$ is a positive integer number. Quite often one refers to this choice of the form factor as the $GF_N$ model.

After solving equation (\ref{GF}) we define the nonlocal potential $\Phi$ as follows
\begin{equation} \n{PPP}
\Phi=-8\pi M \Re(G(X,Y,\zz))\, .
\end{equation}

To find the nonlocal Green function $G(X,Y,\zz)$ we proceed in the same way as in the previous section.
Namely, we use again the Fourier transform in $(X,Y)$ variables
\begin{equation}\n{n1_19}
 G(X,Y,\zz)=\frac{1}{(2\pi)^2}\int e^{-i \vec{\eta}\cdot \vec{\rho}} \tilde{G}(\vec{\eta},\zz) d^2\vec{\eta}\, ,
\end{equation}
and the following representation for the function $\tilde{G}(\vec{\eta},\zz)$
\begin{equation}\n{n1_20}
\tilde{G}(\vec{\eta},\zz)=   \lim_{\sigma \to \infty} \frac{1}{2\pi} \int_{-\infty}^{\infty} e^{-i\zz p}e^{-p^2/2\sigma^2}\tilde{G}(\eta,p)\ dp\, .
\end{equation}
Then using equation (\ref{GF}) one finds
\begin{equation}
\tilde{G}(\eta,p)=-\frac{1}{f(\eta^2+p^2)(\eta^2+p^2)}\, .
\end{equation}
Here $\tilde{G}(\eta,p)$ is the Fourier transform of the Green function \eqref{GF}. It depends on the parameters $\vec{\eta}$ and $p$ of this transform with $\eta^2=\vec{\eta}^2$. It looks quite similar to the expression \eqref{GGPP} with the only difference that now it contains an extra factor $f(\eta^2+p^2)$ in the denominator associated with the form factor.

Combining these results  one gets
\begin{equation}
\begin{split}
    G(\rho,\zz) = &-\frac{1}{(2\pi)^3}\int d^2\eta  e^{-i \vec{\eta}\cdot \vec{\rho}  }\, Y(\eta,\zz)\, ,\\
    Y(\eta,\zz)=&\lim_{\sigma \to \infty} \int_{-\infty}^{\infty}  dp \frac{e^{-p^2/2\sigma^2} e^{-ip \zz}}{f(\eta^2+p^2)(\eta^2+p^2)} \, .
\end{split}
\end{equation}
Using (\ref{BES}) we can write the expression for $G(\rho,\zz)$ in the form
\begin{equation}\n{GG}
 G(\rho,\zz)=-\frac{1}{4\pi^2}\int_0^{\infty} d\eta \eta Y(\eta,\zz) J_0(\eta\rho)\, .
\end{equation}
For the $GF_N$ model the integral in $Y(\eta,\zz)$ contains an exponentially decreasing factor $\sim \exp([-(\ell^2(\eta^2+p^2))^N]$ which provides the convergence of the integral. For this reason, one can  simply put $\sigma=\infty$ in the integrand\footnote{
This remark is valid for any sufficiently fast decreasing at $|p|\to\infty$ form factors.}.

In the simplest case when $N=1$, the form factor takes the form
\begin{equation}
f(\eta^2 + p^2) = e^{\alpha(\eta^2 + p^2)}\hh
\alpha=\ell^2\, ,
\end{equation}
and one has\begin{equation}\n{YY}
Y(\eta,\zz)= 2e^{-\alpha\eta^2}\int_0^{\infty} dp e^{-\alpha p^2 } \frac{\cos(p\zz)}{\eta^2 + p^2} \, .
\end{equation}
For this case, the Green function can be found exactly in an explicit form. In what follows we shall focus on this case.

\subsection{Nonlocal Green function}

Relations (\ref{GG}) and (\ref{YY}) give the required integral representation for the nonlocal Green function. In fact, this function depends on the polar coordinates $\rho$ and $z$, so we write it as $G(\rho,\zz)$. For the $GF_1$ model this Green function can be found in an explicit form.
For this purpose, we use the following relation
\begin{equation}\label{eqI}
    \frac{d}{d\alpha}{Y}  = - A \, ,
\end{equation}
where
\begin{equation}
\begin{split}
    A  &= 2e^{-\alpha\eta^2}\int_0^{\infty} dp e^{-\alpha p^2 } \cos(p\zz)\\
    &=\frac{\sqrt{\pi}}{\sqrt{\alpha}}e^{-\alpha\eta^2}e^{-\zz^2/4\alpha} \, .
\end{split}
\end{equation}

Differentiating \eqref{GG} with respect to $\alpha$ one gets
\begin{equation}
    \frac{dG}{d\alpha}=\frac{1}{4\pi^2}\int_0^{\infty}
    d\eta \eta A J_0(\eta\rho)\, .
\end{equation}
Taking this integral one finds
\begin{equation}\n{HEAT}
    \begin{split}
   \frac{dG}{d\alpha}&=K(\vec{X};\alpha)\, ,\\
   K(\vec{X};\alpha)&=\frac{\exp\left(-\frac{\rho^2+\zz^2}{4\alpha}\right) }{8\pi^{3/2}\alpha^{3/2}} \, .
   \end{split}
\end{equation}
Integration over $\alpha$ and putting $\alpha=\ell^2$ gives
\begin{equation}
    G(\rho,\zz)=-\frac{1}{4\pi}\frac{\erf\left(\sqrt{\rho^2+\zz^2}
    /2\ell\right)}{\sqrt{\rho^2+\zz^2}}\, .
\end{equation}

Let us note that
\begin{equation}
    \rho^2+\zz^2=(r+iy)^2\hh
    y=a\cos(\theta)\, .
\end{equation}
Thus one has
\begin{equation}
G(r,y)=-\frac{1}{4\pi}\frac{\erf\left(\frac{r+iy}{2\ell}\right)}{r+iy}\, .
\end{equation}
In what follows we shall use the following properties of the error function
\begin{equation}
   \erf(-z)=-\erf(z)\hh \overline{\erf(\zeta)}=\erf(\bar{\zeta})\, .
\end{equation}

Let us discuss the properties of the obtained nonlocal Green function. It is a function of the complex variable
\begin{equation}
\zeta=\dfrac{r+iy}{2\ell}\, ,
\end{equation}
and can be written in the form
\begin{equation}
G(r,y)\equiv G(\zeta)=-\frac{1}{8\pi\ell}\frac{\erf(\zeta)}{\zeta}\, ,
\end{equation}

The function $G(\zeta)$ has the following properties
\begin{equation}\n{Gprop}
    G(-\zeta)=G(\zeta)\hh \overline{G(\zeta)}=G(\bar{\zeta})\, .
\end{equation}
The potential $\Phi$ is obtained by taking the real part of $G$. One can write
\begin{equation}
\begin{split}
\Phi&=-4\pi M G_R\, ,\\
G_R(\zeta)&=2Re(G(\zeta))=G(\zeta)+\overline{G(\zeta)}\, .
\end{split}
\end{equation}

In the $\R_+$ domain where   $r>|y|$, the error function remains finite at infinity. For fixed values of $r$ and $y$  one has
\begin{equation}
    \lim_{\ell\to 0}\erf\left(\frac{r+iy}{2\ell}\right)=1\, .
\end{equation}
Thus
\begin{equation}
    \lim_{\ell\to 0}G(r,y)=-\frac{1}{4\pi}\frac{1}{r+iy}\, .
\end{equation}
This means that in the local limit, that is when $\ell\to 0$, the constructed nonlocal Green function correctly reproduces the local Green function (\ref{G0}).

However, this property is violated in $\R_-$ where $r<|y|$. In this domain,  the Green function $G(r,y)$ does not properly reproduce the local Green function in the limit $\ell\to 0$. Let us discuss this point in more detail.

At the boundary surface $\pa\R$ separating the $\R_+$ and $\R_-$ domains one has $r=|y|$. Calculating the value of $G_R(\zeta)$ on $\pa \R$ one gets
\begin{equation}
G_R(\zeta)|_{\pa\R}=G[r(1-i\lambda)]+G[r(1+i\lambda)]\, ,
\end{equation}
where $\lambda=\sgn(y)$. Let us denote
\begin{equation}
\begin{split}
\tilde{G}(\zeta)&=G(i\zeta)\, ,\\
\tilde{G}_R(\zeta)&=\tilde{G}(\zeta)+\overline{\tilde{G}(\zeta)}\, .
\end{split}
\end{equation}
Using (\ref{Gprop}) it is easy to check that the value of $\tilde{G}_R(\zeta)$ restricted to the sphere $\pa\R$ coincides with a similar value of $G_R(\zeta)$
\begin{equation}
G_R(\zeta)\big|_{\pa\R}=\tilde{G}_R(\zeta)\big|_{\pa\R}\, .
\end{equation}
We use $\tilde{G}(\zeta)$ to define the potential $\Phi$ in the domain $\R_-$.
As a result, we obtain the following expression for the potential $\Phi$ which is valid in both domains $\R_{\pm}$ (see Fig.~\ref{F3})
\begin{equation}\n{POTEN}
\Phi=\mu\,  \Re\left(\dfrac{\erf(\zeta)}{\zeta}\right)\, .
\end{equation}
Here $\mu=M/\ell$ and
\begin{equation}\n{zeta}
\zeta= \begin{cases}
    \dfrac{r+iy}{2\ell}\, , & r>|y| \\
    \dfrac{y+ir}{2\ell}\, , & r<|y|
\end{cases}
\end{equation}
This so-defined potential is continuous at $\pa\R$ and has a correct local limit when $\ell\to 0$.
Using the definition of the complementary error function
\begin{equation}\n{erfc}
\erfc(z)=1-\erf(z)\, ,
\end{equation}
one can write the potential $\Phi$ in the form
\begin{equation}\n{PhiPsi}
\Phi=\Phi_0+\Psi\, .
\end{equation}
Here $\Phi_0$ is the potential for the local theory given by \eqref{P0}
\be
\Phi_0=\mu \Re\left(\dfrac{1}{\zeta}\right)\, ,
\ee
and
\begin{equation}\n{PSI}
\Psi=-\mu \Re\left(\dfrac{\erfc(\zeta)}{\zeta}\right)\, .
\end{equation}
The function $\Psi$ describes the nonlocality contribution to the potential $\Phi$.
The complex variable $\zeta$ is defined by \eqref{zeta}.

\begin{figure}[!hbt]
    \centering
      \includegraphics[width=0.55\textwidth]{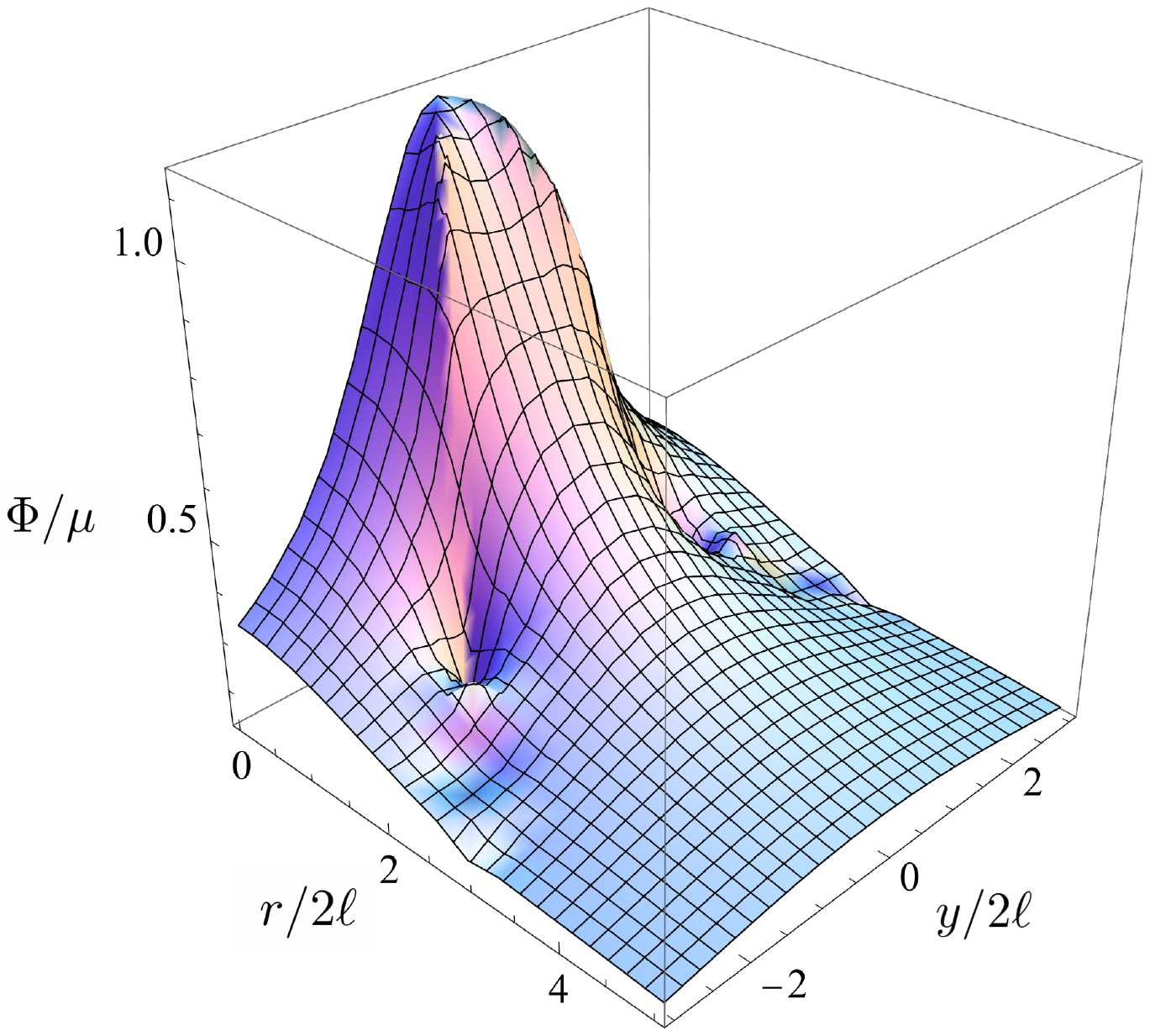}
      \vspace{-0.8cm}
    \caption{\n{F3} Plot of a potential $\Phi/\mu$ as a function of $(r/2\ell,y/2\ell)$.}
\end{figure}

Before we discuss properties of the nonlocal potential $\Phi$ let us make the following remark.
The function $K$ which enters the equation (\ref{HEAT}) has the form
\begin{equation}
\begin{split}
K(\vec{X},\alpha)&=\dfrac{\exp\left(-\frac{\vec{X}^2}{4\alpha}\right) }{{8}\pi^{3/2}\alpha^{3/2}}\, ,\\
\vec{X}^2&=X^2+Y^2+(Z+ia)^2\n{INTER}
\, .
\end{split}
\end{equation}
It is easy to check that this function obeys the following heat equation
\begin{equation}
\dfrac{\pa K}{\pa\alpha}-\lap K=0\, ,
\end{equation}
where $\lap=\pa^2_X+\pa^2_Y+\pa^2_Z$ is the standard flat Laplacian. Thus $K$ can be considered as a heat kernel in a space with the interval $\vec{X}^2$. Let us mention that the method of the heat kernels has been  used earlier for the study of solutions of higher and infinite derivative linearized gravity equations \cite{Frolov:2015bta,Boos:2020,Boos:2020qgg,Kolar:2022kgx}.

The real part of this interval $\vec{X}^2$ is positive in the $\R_+$ domain and negative in the $\R_-$ domain. The problem with the definition of the Green function in $\R_-$ is similar to the problem of defining the heat kernel in the Minkowski space with the Lorentzian signature of the metric. This problem is solved by using the complex parameter $\alpha$ and choosing a proper branch of the corresponding complex function. For more details see e,g. \cite{dewitt1965dynamical,DeWitt:1975ys}.

    \subsection{Properties of the potential}

Let us discuss now some of the properties of the potential $\Phi$ defined by \eqref{POTEN}.

\subsubsection{Potential \texorpdfstring{$\Phi$}{phi} at the ring}

To obtain the value of the potential $\Phi_{ring}$ at the ring, $r=y=0$, it is sufficient to use the following expansion of the error function \cite{abramowitz,oldham2010atlas}
\begin{equation}
\erf(\zeta)=\dfrac{2\zeta}{\sqrt{\pi}}+O(\zeta^2)\, .
\end{equation}
One has
\begin{equation}\n{POT_RING}
\Phi_{ring}=\dfrac{2\mu}{\sqrt{\pi}}\, .
\end{equation}
Hence the potential at the ring is finite and independent of the rotation parameter $a$.

\subsubsection{Potential \texorpdfstring{$\Phi$}{phi} at the symmetry axis}

Let us consider the value of the potential $\Phi$ at the symmetry axis $\theta=0$. For $\theta=\pi$ its value is the same. One has
\begin{equation}\n{AXIS}
\Phi_{axis}=\mu\,  \Re\left(\dfrac{\erf(\zeta)}{\zeta}\right)\, ,
\end{equation}
where
\begin{equation}
\zeta= \begin{cases}
    \dfrac{r+ia}{2\ell}\, , & r>|y| \\
    \dfrac{a+ir}{2\ell}\, , & r<|y|
\end{cases}
\end{equation}
The plot of  $\Phi_{axis}$ is shown in Fig.~\ref{PAXIS}.

\begin{figure}[!hbt]
    \centering
      \includegraphics[width=0.4\textwidth]{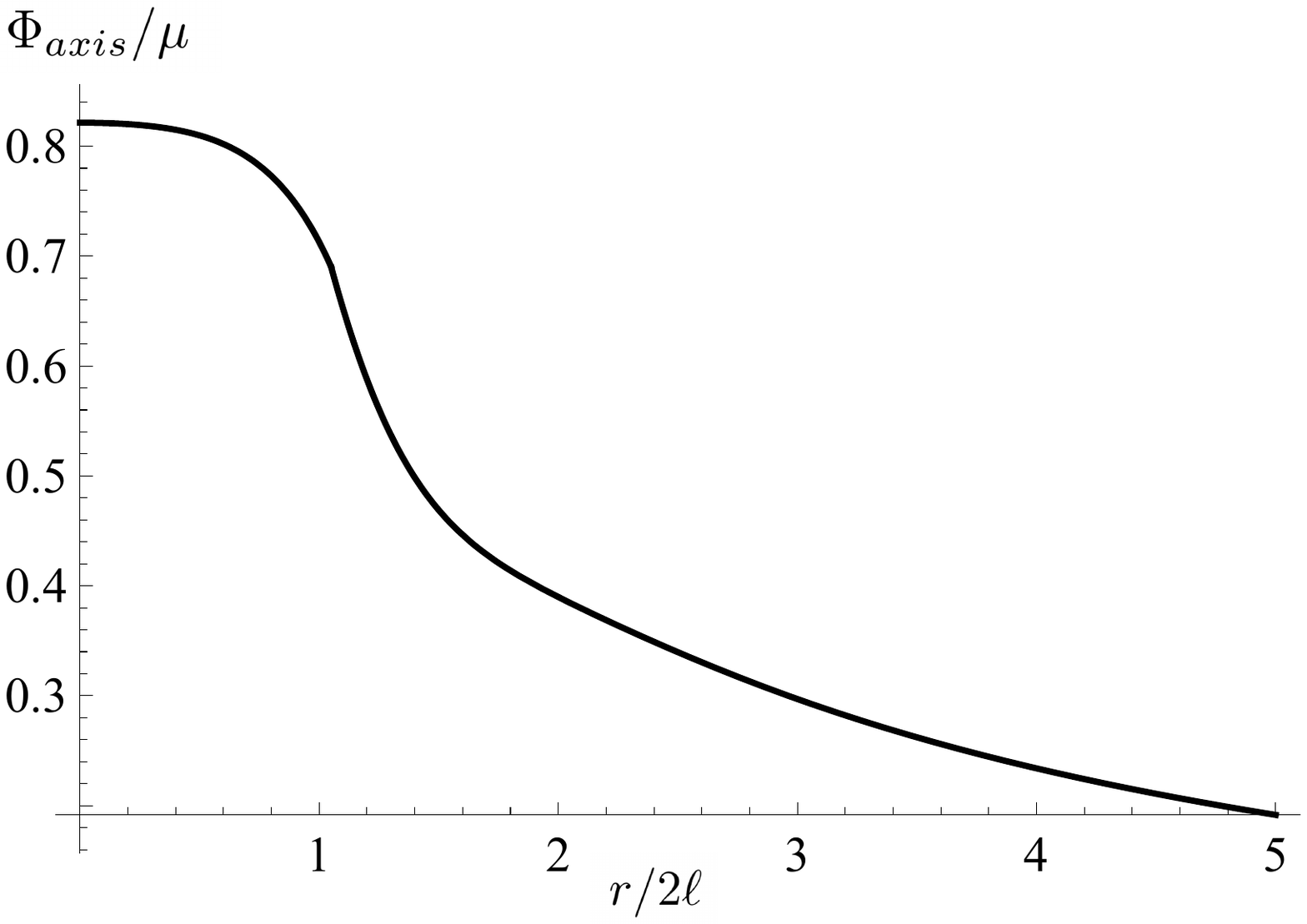}
      %\vspace{-0.8cm}
    \caption{\n{PAXIS} Plot of the potential $\Phi_{axis}$ along the $\theta = 0$ axis
    as a function of $r/2\ell$ for $a/2\ell = 1.05$. }
\end{figure}

\subsubsection{Potential \texorpdfstring{$\Phi$}{phi} on the disc \texorpdfstring{$\mathcal{D}$}{D}}

\begin{figure}[!hbt]
    \centering
      \includegraphics[width=0.4\textwidth]{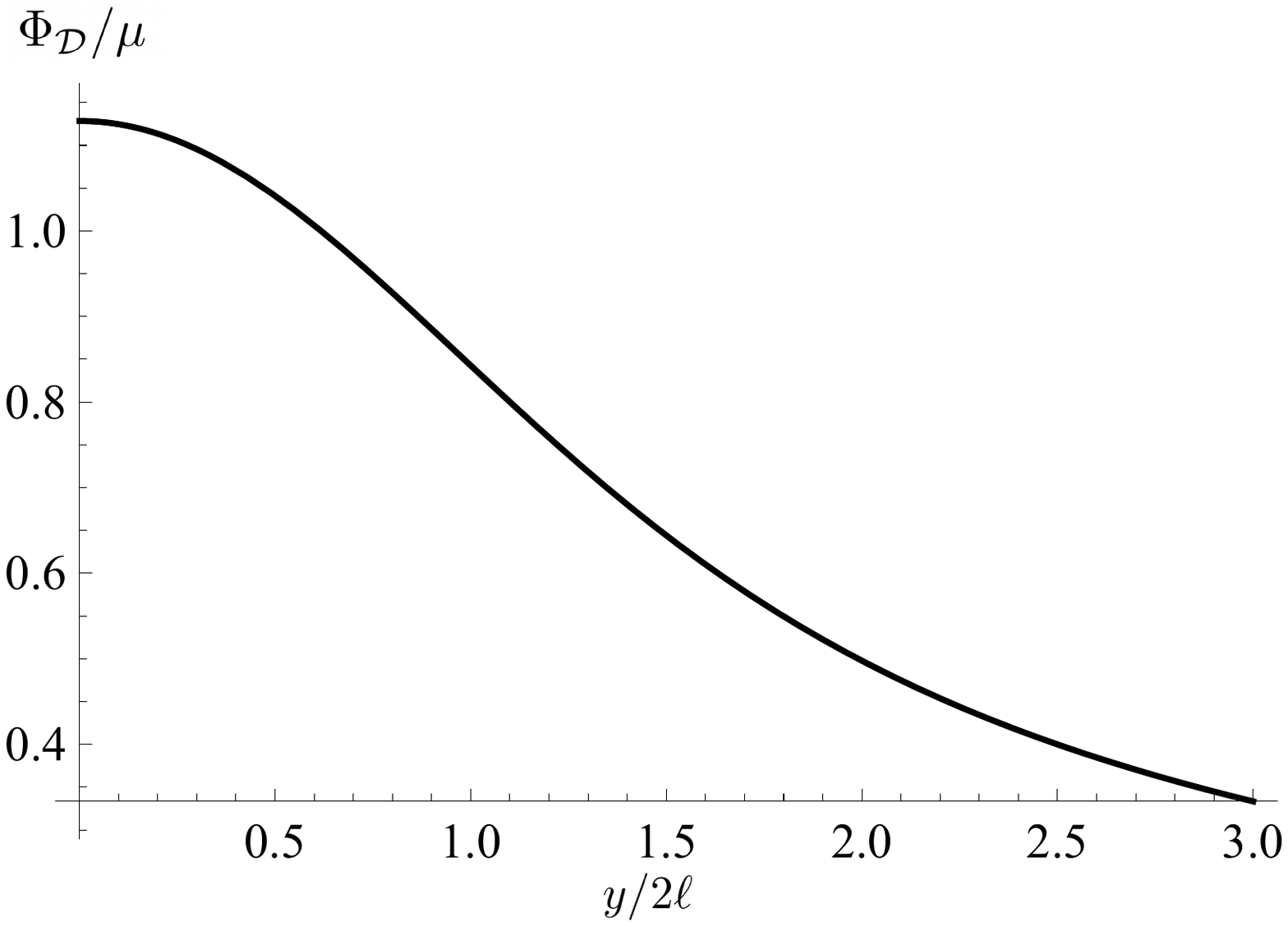}
      %\vspace{-0.8cm}
    \caption{\label{DS} Plot of  the potential  $\Phi_{\mathcal{D}}$  on the disc $\mathcal{D}$ as a function of $\hat{y}=y/2\ell$. }
\end{figure}

The disc  $\mathcal{D}$ is defined by the equation $r=0$, while $0<|y|< a$ and $\phi\in(0,2\pi)$ are the coordinates on the disc. The potential $\Phi$ evaluated on the disc is
\begin{equation}
\Phi_{\mathcal{D}}=\mu\,  \Re \left( \dfrac{\erf(\zeta_0)}{\zeta_0} \right),
\end{equation}
where $\zeta_0=y/(2\ell)$. The plot of  $\Phi_{\mathcal{D}}$ is shown in Fig.~\ref{DS}.

The point $y=0$ corresponds to the ring and the value of $\Phi_{\mathcal{D}}$ at this point coincides with \eqref{POT_RING}. For the disc of the radius $a$ the part of the plot in Fig.~\ref{DS} with $|y|>a$ should be omitted.
At the center of the disc of radius $a$, that is for $y=a$,
the value of  $\Phi_{\mathcal{D}}$ coincides with the limit $r=0$ of the potential $\Phi_{axis}$ on the symmetry axis \eqref{AXIS}.

\subsubsection{Potential \texorpdfstring{$\Phi$}{phi} on the sphere \texorpdfstring{$\pa\R$}{dR}}

\begin{figure}[!hbt]
    \centering
      \includegraphics[width=0.4\textwidth]{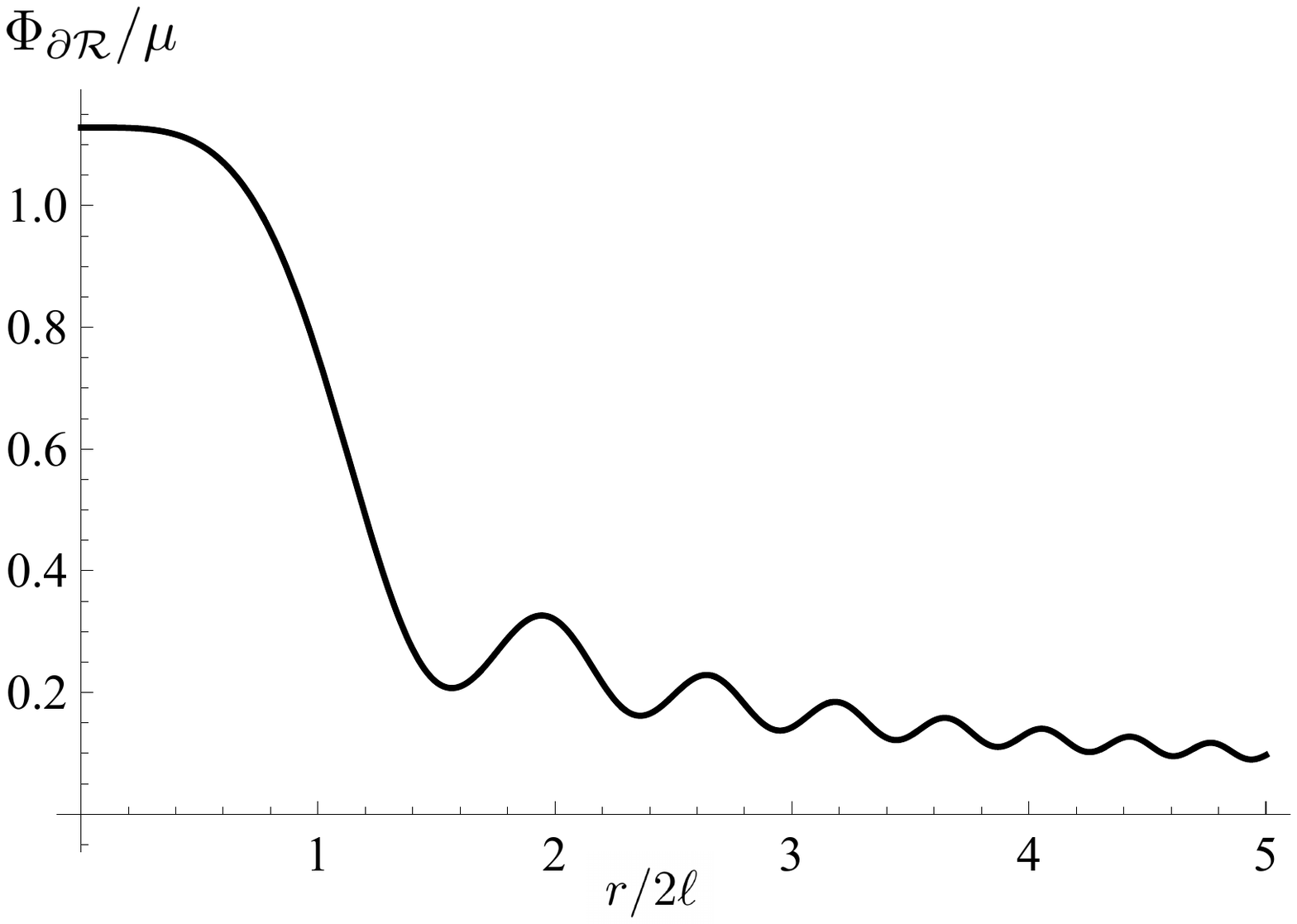}
      \vspace{-0.8cm}
    \caption{\n{RS} Plot of the potential $\Phi$ on the surface $\pa\R$
    as a function of $\hat{r}=r/2\ell$.}
\end{figure}

At the sphere $\pa\R$ one has  $r=|y|$ and the potential $\Phi$ is
\begin{equation}\n{PaR}
\Phi_{\pa\R}=\mu \, \Re \left( \dfrac{\erf(\zeta_0)}{\zeta_0} \right)\hh  \zeta_0=(1+i)\dfrac{r}{2\ell}
\, .
\end{equation}
The plot of  $\Phi_{\pa\R}$ is shown in Fig.~\ref{RS}. For $r=0$, that is, on the ring $\pa\mathcal{D}$, the potential $\Phi_{\pa\R}$ coincides with \eqref{POT_RING}.

\subsubsection{Small \texorpdfstring{$\ell$}{l} limit}

 One can expect that  when $\ell$ is small then $\Psi$ is small as well. Let us discuss this regime in more detail.

For small $\ell$ the argument of the  function $\Psi$ defined by \eqref{PSI} becomes large. In both cases, that is when $r>|y|$  and when $r<|y|$, one can use the following asymptotic form of the complementary error function \cite{abramowitz}
\begin{equation}\n{ASYM}
\erfc(\zeta)=\dfrac{1}{\sqrt{\pi}\zeta}e^{-\zeta^2}+\ldots
\, .
\end{equation}
The nonlocal contribution to the potential $\Psi$  for small $\ell$ is
\begin{equation}\n{SMALL_L}
\Psi(r,y) =-\dfrac{\mu}{\sqrt{\pi}} \Re \left( \dfrac{e^{-\zeta^2}}{\zeta^2}\right)
\, .
\end{equation}

\section{Nonlocal modification of the Kerr metric}
\n{S5}
\subsection{Ergoregion and its inner boundary}

We use the Kerr-Schild ansatz and write the nonlocal modification of the Kerr metric in the form
(\ref{KSP}), where $\Phi$ is the nonlocal potential described in the previous section. Let us notice that the quantity $\Sigma\Phi$ depends not only on the ``radial" coordinate $r$, but also on the ``angle" coordinate $y$. This  difference from the standard (local) Kerr metric has several important consequences
\begin{itemize}
    \item In a general case,  by using transformations similar to (\ref{BL}) one cannot restore the Boyer-Lindquist form of the metric with only one non-vanishing non-diagonal component of the metric $g_{t\phi}$;
    \item The nonlocal version of the metric still has two Killing vectors $\ts{\xi}_{(t)}=\pa_t$ and $\ts{\xi}_{(\phi)}=\pa_{\phi}$, but these vectors do not satisfy the circularity conditions (\ref{CIRC});
    \item As a result of the violation of the circularity conditions, in the general case the surface $V=0$ is not the event horizon.
\end{itemize}

Let us discuss the last point in more detail. The function $V$ vanishes when the following equation is satisfied
\be
\mathcal{V} \equiv \Delta^0_r-\Sigma\Phi=0\, .
\ee
Calculations give
\begin{equation}
\begin{split}
(\nabla \mathcal{V})^2&\equiv \mathcal{V}_{;\mu}\mathcal{V}^{;\mu}=\dfrac{1}{\Sigma} \left[
\Delta^0_y (\Sigma\pa_y\Phi+2y\Phi)^2\right.\\
&\left. +V (\Sigma\pa_r\Phi +2r(\Phi-1))^2\right]\, .
\end{split}
\end{equation}
On the surface  $\mathcal{S}_V$, where $V=0$, the second term in the square brackets vanishes, while the first one is
$\Delta_y^0 \, [\pa_y(\Sigma \Phi)]^2$.
If $\pa_y(\Sigma \Phi)\ne 0$ and $|y|<a$, then $(\nabla \mathcal{V})^2> 0$.
This means that in a general case, the surface $\mathcal{S}_V$ outside the symmetry axis is timelike and hence it cannot be the event horizon.

For the metric \eqref{KSP} a surface $ \mathcal{S}_H$ where $g_{tt}\equiv \ts{\xi}_{(t)}^2=0$ is defined by the relation
\begin{equation}
\Phi=1\, .
\end{equation}
This is an infinite red-shift surface. Outside it, a particle can be at rest with respect to infinity, so that its 4-velocity
\begin{equation}
U^{\mu}=\xi_{(t)}^{\mu}/|\ts{\xi}_{(t)}^2|^{1/2}\, ,
\end{equation}
is timelike.

The domain between $\mathcal{S}_0$ and $\mathcal{S}_V$ is the ergoregion. In this domain, a particle can move along a circular orbit so that its 4 velocity is proportional to a linear combination of the Killing vectors
\begin{equation}
\eta^{\mu}=\xi_{(t)}^{\mu}+\omega \xi_{(\phi)}^{\mu}\, ,
\end{equation}
where $\omega$ is a constant angular velocity. The vector $\ts{\eta}$ is timelike when $\omega\in (\omega_-,\omega_+)$, where
\begin{equation}
\omega_{\pm}=\dfrac{-\ts{\xi}_{(t)}\cdot \ts{\xi}_{(\phi)}\pm \sqrt{V}}{\ts{\xi}_{(\phi)}^2} \, .
\end{equation}
For $\omega=\omega_{\pm}$ the vector $\ts{\eta}$ is null.
At $\mathcal{S}_V$
\begin{equation}
\Omega=\omega_-=\omega_+=-\dfrac{\ts{\xi}_{(t)}\cdot \ts{\xi}_{(\phi)}}{\ts{\xi}_{(\phi)}^2} \, .
\end{equation}
This quantity $\Omega$ is known as the angular velocity of the black hole. We call the surface $\mathcal{S}_V$ the inner boundary of the ergoregion.

In the Kerr metric, the surface $\mathcal{S}_V$ coincides with the horizon and hence is null. It plays the role of a one-way membrane. For the metric \eqref{KSP} with a more general potential function $\Phi$ the situation is quite different. The surface $\mathcal{S}_V$ is timelike, and it can be penetrated by the out-going particles and light rays.

The inner boundary $r=r_V(y)$ of the ergoregion, where $V=0$ is defined by the equation
\begin{equation}
r^2+a^2-2Mr=\Sigma \Psi\, ,
\end{equation}
where $\Psi$ is defined by \eqref{PSI}.
Let us emphasize this relation is valid for an arbitrary function $\Psi$.

For small $\Psi$ the surface $\mathcal{S}_V$ is located close to the unperturbed Kerr horizon,
\be
r=r_H=M+b\hh b=\sqrt{M^2-a^2}\, .
\ee
Let us write
\begin{equation}
h_V(y)=r_V(y)-r_H\, ,
\end{equation}
Then
\begin{equation}\n{hVV}
\hat{h}_V(y)\equiv \dfrac{1}{M}h_V(y)=\left[ \dfrac{\Sigma}{2b}\Psi\right]_{r=r_h}\, .
\end{equation}

For the $GF_1$ model, using the expression \eqref{PSI} for $\Psi$,  one gets
\begin{equation}\n{hVy}
\begin{split}
&\hat{h}_V(y)=-f(x)\, ,\\
&f(x)=\dfrac{\mu}{2\hat{b}}( (1+\hat{b})^2+(1-\hat{b}^2)x^2) \Re\left(\dfrac{\erfc(\zeta)}{\zeta}\right)\, .
\end{split}
\end{equation}
Where we have defined
\begin{equation}\n{VAR_DLESS}
  x=\dfrac{y}{a} \hh \hb=\dfrac{b}{M}\hh \mu=\dfrac{M}{\ell} \, .
\end{equation}

\subsection{Shift of the event horizon}

For a stationary black hole, the event horizon coincides with the outer trapped surface. A useful formalism for finding such surfaces was developed by Senovilla \cite{Senovilla:2011fk}. In this section, we follow this work and apply its results to find the event horizon for the nonlocal modification of the Kerr metric.

Let us assume that in the vicinity of the horizon the potential $\Phi$
differs from its unperturbed (classical) value $\Phi_0$ only slightly.
Hence $\Psi$ defined by \eqref{PhiPsi} is small,
and one can expect that the displacement $h(y)$ of the horizon for the nonlocal modification of the Kerr metric $r_{H,\ell}$  from the Kerr horizon $r_H$ is also small and write
\begin{equation}
    r=r_{H,\ell}\equiv r_H+h(y)\, ,
\end{equation}
where $h(y)$ is small.
At the moment we do not specify the function $\Psi$. We only assume that it is an even function of $y$.
In appendix \ref{A1}, it is shown that the function $h(y)$ obeys the following  linear second order ordinary differential equation which is valid in the leading order of the smallness parameter
\begin{equation}\n{HOR_hh}
\begin{split}
&\dfrac{d}{dy}\left[ (a^2-y^2)\dfrac{dh}{dy}\right]-(\alpha+\tilde{\beta} y^2)h=\varpi\Psi\, ,\\
&\alpha=\dfrac{b}{4M^2r_H^2}\left(
M(4M^2+7Mb+4b^2)+b^3\right)\, ,\\
&\tilde{\beta}=\dfrac{b^2}{4M^2r_H^2}\hh \varpi=-\dfrac{1}{2b} (r_H^2+y^2)(\alpha+\tilde{\beta} y^2)
\, .
\end{split}
\end{equation}
Here $b=\sqrt{M^2-a^2}$.

\subsection{Numerical results}

To find a solution for the horizon shift it is convenient to write the equation \eqref{HOR_hh} in dimensionless form by using $\hat{h} = h/M$, $x=\cos\theta$ and \eqref{VAR_DLESS}
\begin{equation}\n{hhh}
\begin{split}
&\dfrac{d}{dx}\left[ (1-x^2)\dfrac{d\hat{h}}{dx}\right]-(\alpha+\beta x^2)\hat{h}=
F(x)\, ,\\
&\beta = \tilde{\beta} a^2=\dfrac{\hb^2 (1-\hb^2)}{4(1+\hb)^2}\, ,\\
&\alpha=\dfrac{\hb}{4(1+\hb)^2}(4+7\hb+4\hb^2+\hb^3)
\, ,\\
&F=-\dfrac{1}{2\hb}( (1+\hb)^2+(1-\hb^2)\, x^2)
(\alpha+\beta x^2)\Psi
\, .
\end{split}
\end{equation}

Since $\hat{h}$ is an even function of $x$ it satisfies the following condition
\begin{equation}\n{h0}
  \dfrac{d\hat{h}}{dx}\Big|_{x=0}=0\, .
\end{equation}
Both  $\hat{h}(x)$ and $F(x)$ are regular at the symmetry axis $x=\pm 1$ and near it they can be expanded as
\begin{equation}
\begin{split}
  \hat{h}(x)&=\hat{h}_0+\hat{h}_1(1-x^2)+O( (1-x^2)^2)\, ,\\
  F(x)&=F_0+F_1(1-x^2)+O( (1-x^2)^2)\,
\end{split}
\end{equation}
Substituting these expansions in \eqref{hhh} one obtains the following relation
\begin{equation}\n{h1}
 \left[ \dfrac{d\hat{h}}{dx}+\dfrac{1}{4}(\alpha+{\beta})\hat{h}-\dfrac{1}{4}F
 \right]_{x=\pm 1}=0\, .
\end{equation}
Equation \eqref{hhh} with boundary conditions \eqref{h0} and \eqref{h1} is a well posed boundary value problem which can be solved numerically.

Let us first show that for $F=0$ the corresponding homogeneous equation \eqref{hhh} does not have a regular solution. Since this equation is invarian under the reflection $\hat{h}(x)\to -\hat{h}(x)$ it is sufficient to consider only the case when $h(0)>0$. Using the initial condition \eqref{h0} one has
\begin{equation}\n{hom}
\dfrac{d\hat{h}}{dx}=\dfrac{1}{1-x^2}\int_0^x (\alpha+{\beta} x^2) \hat{h}(x) dx\, .
\end{equation}
This relation implies that $\hat{h}(x)$ is a positive monotonically growing function of $x$ and, as a result, $d\hat{h}/dx$ infinitely grows at $x=1$.
\footnote{
Let us note that for $F(x)=0$  equation \eqref{hhh} has a form of the equation for the oblate spheroidal angle functions \cite{flammer}. For a given $\beta$ it has a regular solution only for special values of $\alpha$, which are the eigenvalues of this problem. For an adopted form of the coefficients $\alpha$ and $\beta$ this homogeneous equation has  only trivial regular solution is $\hat{h}(y)=0$.
}

In order to find a numerical solution, it is convenient to use a function $\hat{h}(\theta)$ where $x = \cos(\theta)$ for $\theta = (0,\pi)$. One can write \eqref{hhh} in the following form
\be \n{EQ1}
\dfrac{d^2 \hat{h}}{d\theta^2}+\cot(\theta)\, \dfrac{d \hat{h}}{d\theta}-(\alpha+{\beta} \cos^2\theta^2)\hat{h}=
F(\cos^2\theta)\, .
\ee
We are looking for a solution $\hat{h}$ satisfying the condition
\be\n{BCOND}
\dfrac{d\hat{h}}{d\theta}\Big|_{\theta=\pi/2}=0\, ,
\ee
and which is regular at $\theta=0$ and $\theta=\pi$.

\begin{figure}[!hbt]
    \centering
      \includegraphics[width=0.47
      \textwidth]{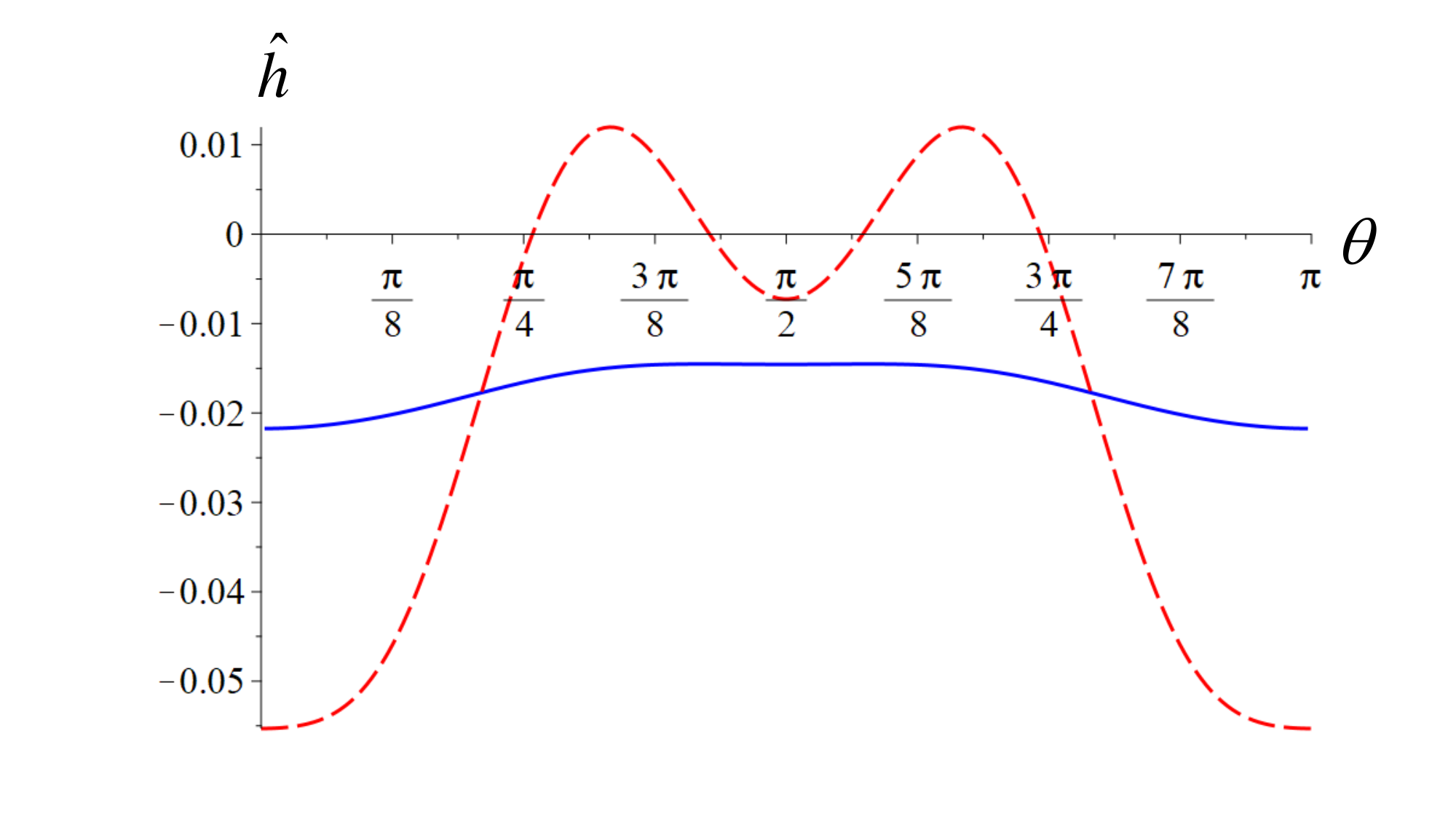}
    \caption{\n{F7}  Plots of $\hat{h}$ (solid line) and $\hat{h}_V(\theta)$ (dash line) as a function of the angle $\theta$ for parameters $\mu=3$ and $\hb=0.3$ ($a=0.9$). }
\end{figure}

\begin{figure}[!hbt]
    \centering
      \includegraphics[width=0.47
      \textwidth]{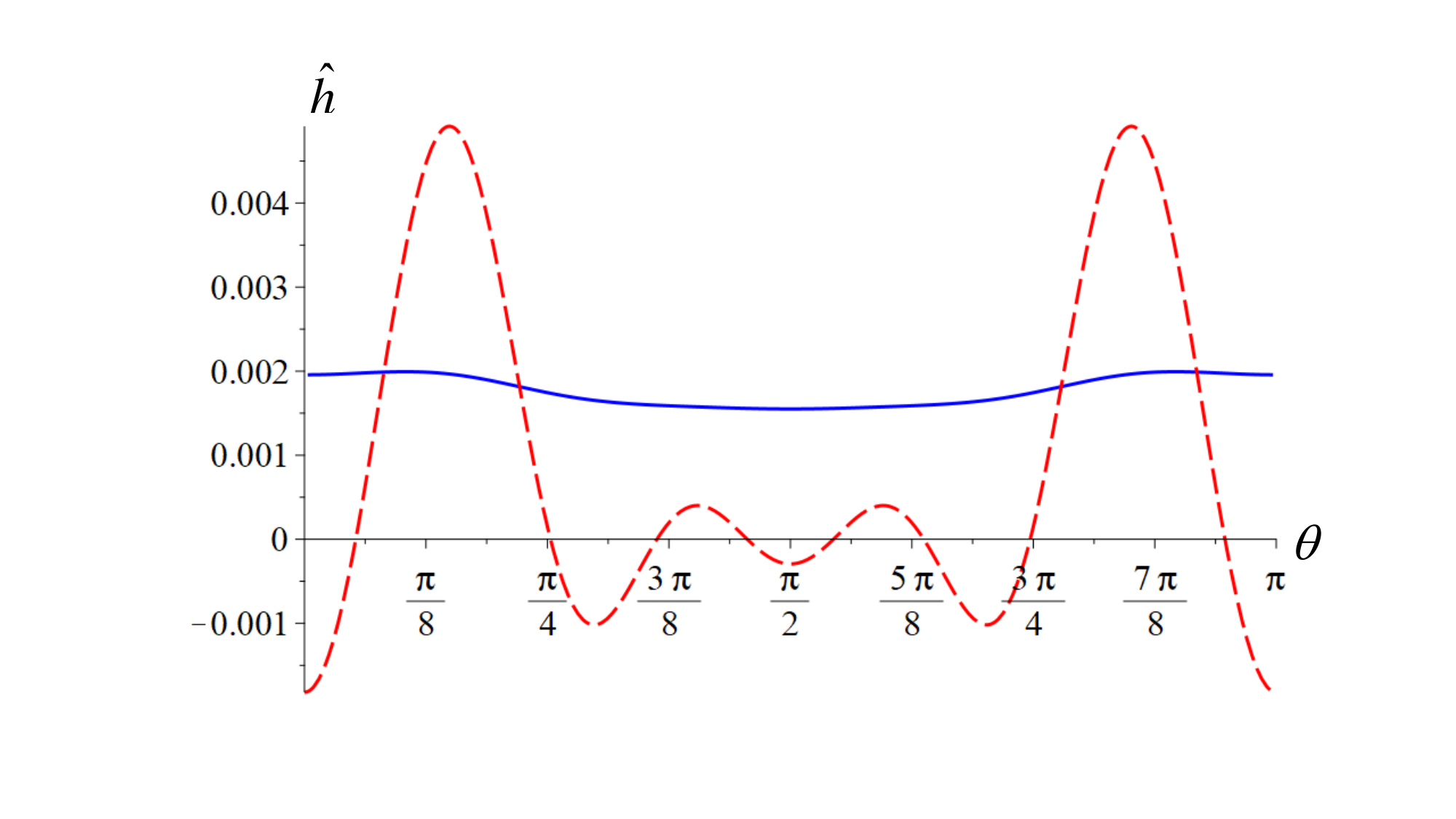}
   \caption{\n{F8} Plots of $\hat{h}$ (solid line) and $\hat{h}_V(\theta)$ (dash line) as a function of the angle $\theta$ for parameters $\mu=4$ and $\hb=0.3$ ($a=0.9$). }
\end{figure}

\begin{figure}[!hbt]
    \centering
      \includegraphics[width=0.53
      \textwidth]{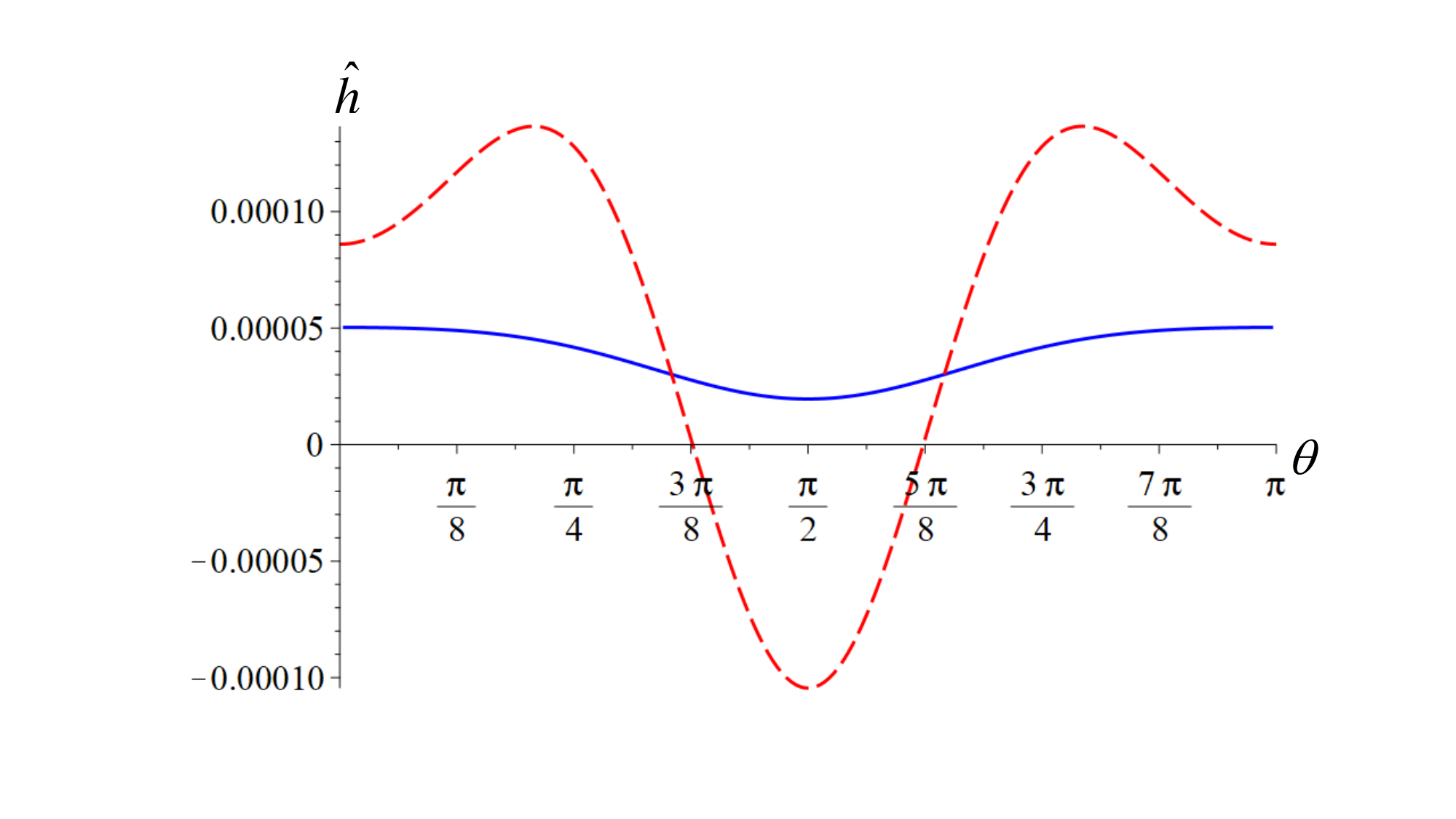}
%\vspace{-3cm}
   \caption{\n{F9} Plots of $\hat{h}$ (solid line) and $\hat{h}_V(\theta)$ (dash line) as a function of the angle $\theta$ for parameters $\mu=3$ and $\hb=0.9$ ($a=0.44$). }
\end{figure}

We chose now the function $\Psi$ in the form \eqref{PSI}. Then the function $F(x)$ which enters the right-hand side of \eqref{hhh} takes the form
\begin{equation}
F(x)=(\alpha+\beta x^2)f(x)\, ,
\end{equation}
where $x=\cos\theta$ and $f(x)$ is given by \eqref{hVy}.
To find a regular solution with the boundary condition \eqref{BCOND} we used a specially designed solver\footnote{This boundary value problem was solved with pseudo-spectral method, with basis functions $b_k = \cos k\theta$ and Gauss collocation grid (corresponding to Type II discrete cosine transform).The authors are grateful to Andrei Frolov for the help.}. Figures~\ref{F7}--\ref{F9} show plots of $h(\theta)$ and $h_V(\theta)$ for some selected values of the parameters $\mu$ and $\hb$.

\section{Non-rotating black holes}
\n{S6}

Let us discuss now the limiting case of a non-rotating black hole when the rotation parameter vanishes. In this case, the metric is spherically symmetric and all the related expressions are greatly simplified.

The metric \eqref{Flat} takes the form
\begin{equation}\label{Flat0}
\begin{split}
d\ox{s}{\!}^2&= - d{t}^2+dh^2\hh dh^2=dr^2+r^2 d\omega^2\, ,\\
d\omega^2&=d\theta^2+ \sin^2\theta d\phi^2 \, .
\end{split}
\end{equation}
This is a flat metric in spherical coordinates.
The null vector $\ts{l}$ has components
\begin{equation}
l_{\mu}dx^{\mu}=-dt+\epsilon dr\hh l^{\mu}=(1,\epsilon,0,0)\, .
\end{equation}
Here $\epsilon=-1$ for the incoming radial null rays, and $\epsilon=1$ for the outgoing ones.
We define a metric
\begin{equation}\n{KS_a0}
d{s}^2=d\ox{s}{\!}^2+\Phi(l_{\mu}dx^{\mu})^2\, .
\end{equation}
Here $\Phi=\Phi(r)$ is some function.
Written in an explicit form, this metric is
\begin{equation}\n{KS_aa}
d{s}^2=-(1-\Phi) dt^2 -2\epsilon \Phi dt dr+(1+\Phi)dr^2+r^2 d\omega^2 \, .
\end{equation}

One can exclude the non-diagonal term $g_{tr}$ of the metric by making the following coordinate transformation
\begin{equation}\n{TIME}
dt=dt_S - \epsilon\dfrac{\Phi}{1-\Phi}dr\, ,
\end{equation}
The metric \eqref{KS_a0} written in the coordinates $(t_S,r,\theta,\phi)$ is
\begin{equation}\n{SCH}
d{s}^2=-(1-\Phi)\,  dt_S^2 +\dfrac{dr^2}{1-\Phi}+r^2 d\omega^2\, .
\end{equation}
Let us note that the time coordinate $t_S$ differs from time $t$ in the metric \eqref{Flat0}. In fact for $\epsilon = - 1$ one has
\begin{equation}\n{TVR}
dt=dv -dr\hh dv=dt_S+ \frac{dr}{1-\Phi}\, .
\end{equation}
Here $v$ is the standard advanced time coordinate. Let us notice that relation \eqref{TVR} is similar to \eqref{BL} for the Kerr metric and coincides with the latter in the absence of the rotation.
The coordinates $(t,r,\theta,\phi)$ which are used in the Kerr-Schild form of the metric \eqref{KS_a0} cover not only the black hole's exterior but also its interior, remaining regular at the horizon.

One can easily recover the Schwarzschild metric by taking the potential $\Phi$ to be a solution of the equation
\begin{equation}
\lap \Phi=-8\pi M \delta^3(\vec{X})\, .
\end{equation}
Here both the Laplacian $\lap$ and delta function $\delta^3(\vec{X})$ are taken in the real flat space with metric $dh^2$. The solution is
\begin{equation}
\Phi_0=\dfrac{2M}{r}\, ,
\end{equation}
so that the metric \eqref{SCH} is nothing but the Schwarzschild metric.

In order to obtain the nonlocal modification of the Schwarzschild metric it is sufficient to choose the potential $\Phi$ to be a solution of the equation
\begin{equation}
f(\lap)\lap \Phi=-8\pi M \delta^3(\vec{X})\, .
\end{equation}
This equation for the nonlocal $GF_N$ models with the form factor of the form \eqref{GF_N} has been studied in several publications. For $N=1$ and $N=2$ the potential $\Phi^{(N)}$ can be found in an explicit analytic form \cite{Boos_2018,Boos:2020qgg}

\begin{equation}\label{Phi_N}
\begin{split}
\Phi^{(1)}&=2M\dfrac{\erf(\frac{r}{2\ell})}{r}\, ,\\
\Phi^{(2)}&= \dfrac{2M}{3\pi \ell} \left[ 3\Gamma\bigg(\frac{5}{4}\bigg)
{}_1F_3\bigg(\frac{1}{4};\frac{1}{2},\frac{3}{4},\frac{5}{4}; \frac{r^4}{16\ell^4}\bigg)\right. \\
&\left. -\frac{r^2}{2\ell^2} \Gamma\bigg(\frac{3}{4}\bigg)
{}_1F_3\bigg(\frac{3}{4};\frac{5}{4},\frac{3}{2},\frac{7}{4}; \frac{r^4}{16\ell^4}\bigg)
\right]
\, .
\end{split}
\end{equation}
Here ${}_aF_b$ is the hypergeometric function \cite{NIST}.

For all $N$ the potentials $\Phi^{(N)}(r)$ are finite at $r=0$ and they have the following asymptotic form \cite{Boos:2020qgg}
\begin{equation}\label{Phi_N_0}
\begin{split}
\Phi^{(N)} & =\varphi^{(N)}_0+\varphi^{(N)}_2 r^2 +O(r^4)\, ,\\
\varphi_0^{(N)} & = \dfrac{2M}{\pi N\ell}\Gamma\bigg(\frac{1}{2N}\bigg)\, ,\\
\varphi_2^{(N)} & = -\dfrac{4M}{3 N\ell^3}\Gamma\bigg(\frac{3}{2N}\bigg)
\, .
\end{split}
\end{equation}
Let us note that for all $GF_N$ models, the coefficients $\varphi_0^{(N)}$ are finite and positive.
For the nonrotating black hole, the inner boundary of the ergosphere coincides with the event horizon  and its equation is $\Phi=1$. For the $GF_1$ model this  equation can be written in the form
\begin{equation}
\mu \erf(x)=x\hh r=2\ell x
\, .
\end{equation}

\section{Discussion}
\n{S7}

In this paper, we discussed the nonlocal modification of the Kerr geometry. Our starting point is the Kerr-Schild form of the Kerr metric. The potential which enters this representation is a solution of the 3D flat Poisson equation with a point-like source shifted to the complex space. We considered a modification of this equation obtained by changing the Laplace operator $\lap$ by its infinite derivative analog $f(\lap)\lap$.
The function $f(z)$ is chosen so that it does not have zeroes in the complex plane $z$, so that the form factor operator has an inverse. We focus on the study of the simplest case, namely when the form factor has the form $f=\exp(-\ell^2\lap)$. In this case, the potential $\Phi$ can be obtained in an explicit analytic form. We discussed the properties of a rotating black hole in such a nonlocal model.

Let us notice, that in order to reconstruct the Kerr metric in Boyer-Lindquist coordinates, one should make a coordinate transformation that contains dependence on the black-hole's mass $M$. As a result, this parameter enters the Kerr metric in the Boyer-Lindquist coordinates nonlinearly. It is easy to check that a simple linearization of the Kerr metric, by expanding it in terms of the mass parameter and keeping only its zero and first order in $M$ terms, produces a metric that is singular and  does not have a horizon. One can also check that the nonlocal modification of the Kerr metric presented in this paper, like the Kerr metric, is regular at the horizon.

The main difference of the nonlocal modification of the Kerr metric discussed in this paper is that besides the mass $M$ and the rotation parameter $a$ which specify the Kerr solution it contains a new parameter $\ell$ which controls the nonlocality effects. We did not specify its value. However, recent experiments showed that
Newtonian gravity gave an excellent fit to the data at least up to the length  $\ell_{Newton}=38.6 \mu m$ \cite{Lee:2020zjt}. This means that $\ell$ at least should be less than $\ell_{Newton}$. This implies that for astrophysical stellar mass and supermassive black holes $\ell/M\ll 1$.  One can expect that the corresponding nonlocal effects for these objects are extremely small and exponentially suppressed. The effects of the nonlocality discussed in this paper might be important when $\ell/M\sim 1$, that is for mini black holes. In particular, the nonlocality may change the properties of their Hawking evaporation, such as its temperature and anisotropy. One can also expect that the effects of the nonlocality becomes important at the final stage of the mini black hole evaporation.

An important property of the Kerr-Schild form of the Kerr metric is that there exists a coordinate transformation that allows one to recover the Kerr metric which has only one non-diagonal component, $g_{t\phi}$. This property is not valid for the nonlocal modification of the Kerr metric discussed in this paper. This property makes this metric quite different from models of a regular rotating black hole discussed in the papers \cite{Gurses:1975vu,Babichev:2020qpr,torres2022regular,Baines:2023dhq,Zhou:2023lwc}.

The modified metric described in this paper still has two commuting Killing vectors. However, these vectors do not satisfy the circularity condition which plays an important role in prove the uniqueness theorems for the rotating black hole solutions of the Einstein equations. One of the interesting consequences of the violation of the circularity condition is that the event horizon does not coincide with the inner boundary of the ergoregion,  where the invariant $V$, \eqref{VVV}, constructed from the Killing vectors,  vanishes.

When the ``fundamental length" parameter $\ell$, that defines the scale of nonlocality, tends to zero, the obtained nonlocal potential $\Phi$ has the limit $\Phi_0=2Mr/(r^2+y^2)$, and the metric takes the form of the standard Kerr metric. Corrections to the metric in the black hole exterior are controlled by the dimensionless parameter $\ell/M$. When this parameter is small the event horizon of the nonlocal black hole is slightly shifted from the Kerr horizon. In this approximation,  we derived and numerically solved the equation that describes this shift. These results are illustrated by figures \ref{F7}--\ref{F9}. Solid and dashed lines represent the deviation of the modified event horizon and the position of the inner boundary of the ergoregion with respect to the Kerr horizon.

In the absence of the rotation, that is in the limit $a\to 0$, the modified metric contains two parameters, the mass $M$ and the scale of the nonlocality $\ell$. This metric and its properties are discussed in section~VI.
Let us emphasize, that in the Kerr-Schild representation the potential $\Phi$ enters as a perturbation of the flat metric and it is a solution of the linearized infinite derivative gravity equations. The standard "Schwarzschild" type form of the metric \eqref{SCH} is obtained after making the coordinate transformation
\eqref{TVR} which depends on the mass parameter in the nonlinear form.

Let us emphasize that the obtained nonlocal Kerr metric is not a solution to the fundamental nonlocal gravity equations. However, one can expect that it might properly reproduce some important features of the (unknown at the moment) solution for a rotating black hole in the consistent nonlocal (infinity derivative) models of gravity.

\acknowledgments

This work was supported  by  the Natural Sciences and Engineering
Research Council of Canada. The authors are also grateful to the
Killam Trust for its financial support. The authors thank Andrei Frolov for his help with finding the numerical solutions of the equation for the horizon shift.

\appendix

\section{Marginally trapped surface}

\n{A1}

The explicit form of the metric \eqref{KSP} in $(t,r,y,\phi)$ coordinates for an arbitrary function $\Phi = \Phi(r,y)$ is
\begin{equation}\label{g_dn}
    \begin{split}
     d\tilde{s}^2 &= -(1-\Phi)dt^2 - \epsilon\frac{2\Phi \Sigma }{\Delta^0_r}dtdr + \frac{2\Phi \Delta^0_y}{a}dtd\phi \\
     & + \frac{\Sigma(\Phi\Sigma + \Delta^0_r)}{(\Delta^0_r)^2}dr^2 - \epsilon\frac{2\Phi\Sigma\Delta^0_y}{a\Delta^0_r}drd\phi \\
     & + \frac{\Sigma}{\Delta^0_y}dy^2 + (\Delta^0_y\Phi + \Delta^0_r)\frac{\Delta^0_y}{a^2}d\phi^2 \, .
    \end{split}
\end{equation}
The contravariant components of this metric are
\begin{equation}\label{g_up}
    g^{\mu\nu} =
    \begin{pmatrix}
       -1-\Phi & -\epsilon\Phi & 0 &\frac{a\Phi}{\Delta^0_r} \\
       -\epsilon\Phi & \frac{\Delta^0_r - \Phi\Sigma}{\Sigma} & 0 & \epsilon \frac{a\Phi}{\Delta^0_r} \\
       0 & 0 & \frac{\Delta^0_y}{\Sigma} & 0 \\
       \frac{a\Phi}{\Delta^0_r} & \epsilon\frac{a\Phi}{\Delta^0_r} & 0 & \frac{a^2(\Delta^0_r - \Phi\Delta^0_y)}{(\Delta^0_r)^2\Delta^0_y}
    \end{pmatrix} \, .
\end{equation}

To find the event horizon in this metric we follow the recipe described by Senovilla \cite{Senovilla:2011fk}. Because of the symmetry of the metric \eqref{KSP} the horizon surface equation can be written in the form
\begin{equation}
r=F(y)\, .
\end{equation}
Denote
\begin{equation}\n{XX}
x=r-F(y)\, ,
\end{equation}
and consider a set of 2D surfaces $\mathcal{S}$
\begin{equation}\n{SURF}
t=t_0\hh x=x_0\, ,
\end{equation}
where $t_0$ and $x_0$ are constant parameters. A 2D surface $ \mathcal{S}_H$ with $x=0$ is the intersection of the event horizon $\mathcal{H}$ by the 3D surface $t=t_0$. This implies that $ \mathcal{S}_H$ is a marginally trapped surface. To find the function $F(y)$ which determines $ \mathcal{S}_H$ we proceed as follows.

First, we change to the $(t,x,y,\phi)$ coordinates by using the relations
\begin{equation}
dr=dx+f(y)dy\hh f(y)=\dfrac{dF}{dy}\, ,
\end{equation}
and then present the metric \eqref{g_dn} in the form
\begin{equation}\n{MET}
ds^2=g_{ab}dx^a dx^b+2g_{a A}dx^a dx^A+g_{AB}dx^A dx^B\, .
\end{equation}
Indices $a,b$ take values $0,1$ while $A,B$ stand for $2,3$, and we denote
\begin{equation}\n{CC}
x^0=t\hhh x^1=x\hhh x^2=y\hhh x^3=\phi\, .
\end{equation}
The condition that the coordinates $x^a$ are constant specifies a 2D surface
$\mathcal{S}$, with $x^A$ coordinates on it. The metric \eqref{MET} in these new coordinates is
\begin{equation}\n{METR_X}
    \begin{split}
        &g_{a b}dx^adx^b = -(1 + \Phi)dt^2 + \frac{2\Phi \Sigma }{\Delta^0_r}dtdx \\
        & + \frac{\Sigma(\Phi\Sigma + \Delta^0_r)}{(\Delta^0_r)^2}dx^2 \, , \\
        &g_{a A}dx^adx^A = \frac{\Phi\Sigma f}{\Delta^0_r}dtdy + \frac{\Phi \Delta^0_y}{a}dtd\phi \\
        & + \frac{\Sigma f(\Sigma \Phi + \Delta^0_r)}{(\Delta^0_r)^2}dxdy + \frac{\Phi \Delta^0_y \Sigma}{a \Delta^0_r}dxd\phi \, ,\\
        &g_{A B}dx^Adx^B = \\
        &\frac{\Sigma((\Delta^0_r)^2 + f^2\Phi\Sigma\Delta^0_y + f^2 \Delta^0_y\Delta^0_r)}{\Delta^0_y(\Delta^0_r)^2}dy^2 \\
        & + \frac{2 \Phi \Sigma \Delta^0_y f}{a\Delta^0_r}dyd\phi + \frac{\Delta^0_y(\Delta^0_y\Phi + \Delta^0_r)}{a^2}d\phi^2 \, .
    \end{split}
\end{equation}

Let us denote by $\gamma_{AB}$ a two dimensional metric on $\mathcal{S}$ and by $\gamma^{AB}$ its inverse. Following \cite{Senovilla:2011fk} we also introduce the following objects
\begin{equation}\n{SEN}
\begin{split}
G&=\sqrt{\det{g_{AB}}}\equiv e^U\, ,\\
\vec{g}_a&=g_{aA} dx^A\, ,\\
\mbox{div}\, \vec{g}_a&=\dfrac{1}{G}\left( G\gamma^{AB}g_{a A}\right)_{,B}\, ,\\
H_{\mu}&=\delta^{\mu}_a(U_{,a}-\mbox{div}\, \vec{g}_a)
 \, .
\end{split}
\end{equation}

A necessary condition for a 2D surface $\mathcal{S}$ to be marginally trapped is that $\kappa=0$ \cite{Senovilla:2011fk}, where
\begin{equation}
\kappa=-g^{ab}H_{a}H_{b}|_{\mathcal{S}}\, .
\end{equation}

Using the GRTensor package in Maple we calculated $\kappa$ for the metric \eqref{MET}, \eqref{METR_X} with an arbitrary potential function $\Phi(r,y)$. However, the obtained expression is rather long, so we do not reproduce it here.

Instead of this, we consider an approximation where the potential $\Phi$ is close to its local limit
\begin{equation}\n{PHI00}
\Phi_0=\dfrac{2Mr}{r^2+y^2}\, .
\end{equation}
In this case, the horizon surface differs only slightly from the Kerr horizon
\begin{equation}
r=r_H=M+\sqrt{M^2-a^2}\, .
\end{equation}

We denote
\begin{equation}\n{HOR}
\begin{split}
F(y)&=r_H+\lambda h(y)\hh f(y)=\lambda \dfrac{dh}{dy} \, ,\\
\Phi(r,y)&=\Phi_0+\lambda \Psi(r,y)
\, ,
\end{split}
\end{equation}
where we have introduced a dimensionless parameter $\lambda$
which we assume to be small. This parameter is used to control the order of ``smallness" of the different terms that enter the equations. We restrict our calculations by keeping the zero and first order expressions in the decomposition over $\lambda$.
At the end of the calculations, we put $\lambda=1$. For simplicity purposes, we proceed as follows: First, we omit in the metric coefficients all of the terms which contain $f^2$, $f\pa_{\mu}f$ and other similar expressions, which are evidently of second order in $\lambda$. After calculating the quantity $\kappa$ for an arbitrary $\Phi$ we use \eqref{HOR} and omit all of the $O(\lambda^2)$ terms in the final expression.

In the adopted approximation, after omitting quadratic in $f$ terms, one obtains the following expression for the $g_{A B}$ part of the metric \eqref{MET}
\begin{equation}\n{KSP0}
\begin{split}
g_{AB}dx^A dx^B &= \frac{\Sigma}{\Delta^0_y}dy^2 + \frac{2 \Phi \Sigma \Delta^0_y f}{a\Delta^0_r}dyd\phi + \frac{\Delta^0_y\Upsilon}{a^2}d\phi^2 \, .\\
 \Upsilon &= \Delta^0_y\Phi+\Delta^0_r \, ,
\end{split}
\end{equation}
and one has
\begin{equation}
\begin{split}
G&\equiv\sqrt{\det{g_{AB}}}=\dfrac{\sqrt{\Sigma \Upsilon}}{a} \, .
\end{split}
\end{equation}

Let us note that the metric coefficients in \eqref{g_dn} and \eqref{METR_X} are functions of $(r,y)$ coordinates. In order to calculate their partial derivatives with respect to $(x,y)$ variables one should use the relations
\begin{equation}\n{X_DER}
\begin{split}
&\dfrac{\pa B(r,y)}{\pa x}\Big|_{y}=\dfrac{\pa B(r,y)}{\pa r}\Big|_{y}\, ,\\
&\dfrac{\pa B(r,y)}{\pa y}\Big|_{x}=\dfrac{\pa B(r,y)}{\pa y}\Big|_{r}+f\dfrac{\partial B(r,y)}{\pa r}\Big|_{y}
\, .
\end{split}
\end{equation}

The $t-$component of $U_{,a}$ vanishes, while the other component is
\begin{equation}\n{Ux}
U_{,x}= \frac{\pa_r(\Sigma\Upsilon)}{2\Sigma\Upsilon}
\, .
\end{equation}
One also gets
\begin{equation}\n{KSPt}
\begin{split}
&\mbox{div}\vec{g}_t= \frac{1}{2\Sigma\Upsilon^2}\big[\Delta^0_y\Phi \Upsilon(2\Sigma\partial_y f + f \partial_y\Sigma) \\
&+ \Sigma f (\Upsilon + \Delta^0_r)
\pa_y\Upsilon \big] \, ,
\end{split}
\end{equation}
\begin{equation}\n{KSPx}
\begin{split}
&\mbox{div}\vec{g}_x= \frac{1}{2\Sigma\Delta^0_r\Upsilon^2}\big[ f\Delta^0_y \Upsilon(\Upsilon + 3\Sigma\Phi)\partial_y\Sigma \\
& + \Sigma f \Delta^0_y(\Delta_r^0(\Upsilon+\Sigma))\partial_y\Phi \\
& + \Sigma f(\Sigma\Phi(\Upsilon + \Delta^0_r)+\Upsilon(2\Upsilon + \Delta^0_y\Phi))\partial_y\Delta^0_y \\
& + 2\Sigma\Delta^0_y\Upsilon(\Upsilon + \Sigma\Phi)\partial_y  f \big] \, .
\end{split}
\end{equation}

After ubstituting these expressions in $H_{\mu}$ defined by \eqref{SEN} we calculated the quantity $\kappa$. In these calculations
we use the following truncated version of $g^{ab}$ in which only the zero and first order in $f$ is preserved
\begin{equation}
g^{ab}=
  \begin{pmatrix}
      -1-\Phi & \Phi & 0 &\frac{a\Phi}{\Delta^0_r} \\
       \Phi & \frac{\Delta^0_r - \Phi\Sigma}{\Sigma} & -\frac{f\Delta_y^0}{\Sigma} & -\frac{a\Phi}{\Delta^0_r} \\
       0 & -\frac{f\Delta_y^0}{\Sigma} & \frac{\Delta^0_y}{\Sigma} & 0 \\
       \frac{a\Phi}{\Delta^0_r} & -\frac{a\Phi}{\Delta^0_r} & 0 & \frac{a^2(\Delta^0_r - \Phi\Delta^0_y)}{(\Delta^0_r)^2\Delta^0_y}
    \end{pmatrix} \, .
\end{equation}

Following our approximation, we use again relations \eqref{HOR} in the obtained expression for $\kappa$ while retaining solely the leading-order terms with respect to $\lambda$. In particular, this means that it is sufficient to use the quantity $\Psi(r_h,y)$ instead of $\Psi(r,y)$ since $\Psi$ itself is already is of the first order in $\lambda$.

As it is expected, the contribution to $\kappa$ of the order $\lambda^0$ vanishes since $r=r_h$ is the horizon of the unperturbed Kerr metric. In the first order in $\lambda$ the condition $\kappa =0$ gives the following differential equation for the function $h(y)$ which describes the displacement of the horizon for the perturbed metric.

\begin{equation}\n{HOR_h}
\begin{split}
&\dfrac{d}{dy}\left[ (a^2-y^2)\dfrac{dh}{dy}\right]-(\alpha+\tilde{\beta} y^2)h=\varpi\Psi\, ,\\
&\alpha=\dfrac{b}{4M^2r_H^2}\left(
M(4M^2+7Mb+4b^2)+b^3\right)\, ,\\
&\tilde{\beta}=\dfrac{b^2}{4M^2r_H^2}\, ,\\
&\varpi=-\dfrac{1}{2b} (r_H^2+y^2)(\alpha+\tilde{\beta} y^2)
\, .
\end{split}
\end{equation}
Here $b=\sqrt{M^2-a^2}$.

\vspace{2cm}

\bibliography{NON_LOCAL_KERR}

%merlin.mbs apsrev4-1.bst 2010-07-25 4.21a (PWD, AO, DPC) hacked
%Control: key (0)
%Control: author (0) dotless jnrlst
%Control: editor formatted (1) identically to author
%Control: production of article title (0) allowed
%Control: page (1) range
%Control: year (0) verbatim
%Control: production of eprint (0) enabled
\begin{thebibliography}{74}%
\makeatletter
\providecommand \@ifxundefined [1]{%
 \@ifx{#1\undefined}
}%
\providecommand \@ifnum [1]{%
 \ifnum #1\expandafter \@firstoftwo
 \else \expandafter \@secondoftwo
 \fi
}%
\providecommand \@ifx [1]{%
 \ifx #1\expandafter \@firstoftwo
 \else \expandafter \@secondoftwo
 \fi
}%
\providecommand \natexlab [1]{#1}%
\providecommand \enquote  [1]{``#1''}%
\providecommand \bibnamefont  [1]{#1}%
\providecommand \bibfnamefont [1]{#1}%
\providecommand \citenamefont [1]{#1}%
\providecommand \href@noop [0]{\@secondoftwo}%
\providecommand \href [0]{\begingroup \@sanitize@url \@href}%
\providecommand \@href[1]{\@@startlink{#1}\@@href}%
\providecommand \@@href[1]{\endgroup#1\@@endlink}%
\providecommand \@sanitize@url [0]{\catcode `\\12\catcode `\$12\catcode `\&12\catcode `\#12\catcode `\^12\catcode `\_12\catcode `\%12\relax}%
\providecommand \@@startlink[1]{}%
\providecommand \@@endlink[0]{}%
\providecommand \url  [0]{\begingroup\@sanitize@url \@url }%
\providecommand \@url [1]{\endgroup\@href {#1}{\urlprefix }}%
\providecommand \urlprefix  [0]{URL }%
\providecommand \Eprint [0]{\href }%
\providecommand \doibase [0]{http://dx.doi.org/}%
\providecommand \selectlanguage [0]{\@gobble}%
\providecommand \bibinfo  [0]{\@secondoftwo}%
\providecommand \bibfield  [0]{\@secondoftwo}%
\providecommand \translation [1]{[#1]}%
\providecommand \BibitemOpen [0]{}%
\providecommand \bibitemStop [0]{}%
\providecommand \bibitemNoStop [0]{.\EOS\space}%
\providecommand \EOS [0]{\spacefactor3000\relax}%
\providecommand \BibitemShut  [1]{\csname bibitem#1\endcsname}%
\let\auto@bib@innerbib\@empty
%</preamble>
\bibitem [{\citenamefont {Kerr}(1963)}]{Kerr}%
  \BibitemOpen
  \bibfield  {author} {\bibinfo {author} {\bibfnamefont {Roy~P.}\ \bibnamefont {Kerr}},\ }\bibfield  {title} {\enquote {\bibinfo {title} {Gravitational field of a spinning mass as an example of algebraically special metrics},}\ }\href {\doibase 10.1103/PhysRevLett.11.237} {\bibfield  {journal} {\bibinfo  {journal} {Phys. Rev. Lett.}\ }\textbf {\bibinfo {volume} {11}},\ \bibinfo {pages} {237--238} (\bibinfo {year} {1963})}\BibitemShut {NoStop}%
\bibitem [{\citenamefont {Carter}(1973)}]{Carter:1973rla}%
  \BibitemOpen
  \bibfield  {author} {\bibinfo {author} {\bibfnamefont {B.}~\bibnamefont {Carter}},\ }\bibfield  {title} {\enquote {\bibinfo {title} {{Black holes equilibrium states}},}\ }in\ \href@noop {} {\emph {\bibinfo {booktitle} {{Les Houches Summer School of Theoretical Physics}: {Black Holes}}}}\ (\bibinfo {year} {1973})\ pp.\ \bibinfo {pages} {57--214}\BibitemShut {NoStop}%
\bibitem [{\citenamefont {Misner}\ \emph {et~al.}(1973)\citenamefont {Misner}, \citenamefont {Thorne},\ and\ \citenamefont {Wheeler}}]{MTW}%
  \BibitemOpen
  \bibfield  {author} {\bibinfo {author} {\bibfnamefont {C.~W.}\ \bibnamefont {Misner}}, \bibinfo {author} {\bibfnamefont {K.~S.}\ \bibnamefont {Thorne}}, \ and\ \bibinfo {author} {\bibfnamefont {J.~A.}\ \bibnamefont {Wheeler}},\ }\href@noop {} {\emph {\bibinfo {title} {{Gravitation}}}}\ (\bibinfo  {publisher} {W. H. Freeman},\ \bibinfo {address} {San Francisco},\ \bibinfo {year} {1973})\BibitemShut {NoStop}%
\bibitem [{\citenamefont {Chandrasekhar}(1992)}]{chandrasekhar}%
  \BibitemOpen
  \bibfield  {author} {\bibinfo {author} {\bibfnamefont {S.}~\bibnamefont {Chandrasekhar}},\ }\href {https://books.google.ca/books?id=mLMYAQAAMAAJ} {\emph {\bibinfo {title} {The Mathematical Theory of Black Holes}}},\ International series of monographs on physics\ (\bibinfo  {publisher} {Oxford University Press},\ \bibinfo {year} {1992})\BibitemShut {NoStop}%
\bibitem [{\citenamefont {Frolov}\ and\ \citenamefont {Novikov}(1998)}]{Frolov:1998wf}%
  \BibitemOpen
  \bibfield  {author} {\bibinfo {author} {\bibfnamefont {V.P.}\ \bibnamefont {Frolov}}\ and\ \bibinfo {author} {\bibfnamefont {I.D.}\ \bibnamefont {Novikov}},\ }\href@noop {} {\emph {\bibinfo {title} {{Black hole physics: Basic concepts and new developments}}}}\ (\bibinfo  {publisher} {Kluwer Acad. Publ.},\ \bibinfo {year} {1998})\BibitemShut {NoStop}%
\bibitem [{\citenamefont {Stephani}\ \emph {et~al.}(2003)\citenamefont {Stephani}, \citenamefont {Kramer}, \citenamefont {MacCallum}, \citenamefont {Hoenselaers},\ and\ \citenamefont {Herlt}}]{kramer}%
  \BibitemOpen
  \bibfield  {author} {\bibinfo {author} {\bibfnamefont {Hans}\ \bibnamefont {Stephani}}, \bibinfo {author} {\bibfnamefont {Dietrich}\ \bibnamefont {Kramer}}, \bibinfo {author} {\bibfnamefont {Malcolm}\ \bibnamefont {MacCallum}}, \bibinfo {author} {\bibfnamefont {Cornelius}\ \bibnamefont {Hoenselaers}}, \ and\ \bibinfo {author} {\bibfnamefont {Eduard}\ \bibnamefont {Herlt}},\ }\href {\doibase 10.1017/CBO9780511535185} {\emph {\bibinfo {title} {Exact Solutions of Einstein's Field Equations}}},\ Cambridge Monographs on Mathematical Physics\ (\bibinfo  {publisher} {Cambridge University Press},\ \bibinfo {year} {2003})\BibitemShut {NoStop}%
\bibitem [{\citenamefont {O'Neill}(2014)}]{Neill}%
  \BibitemOpen
  \bibfield  {author} {\bibinfo {author} {\bibfnamefont {B.}~\bibnamefont {O'Neill}},\ }\href {https://books.google.ca/books?id=jHXCAgAAQBAJ} {\emph {\bibinfo {title} {The Geometry of Kerr Black Holes}}},\ Dover Books on Physics\ (\bibinfo  {publisher} {Dover Publications},\ \bibinfo {year} {2014})\BibitemShut {NoStop}%
\bibitem [{\citenamefont {Penrose}(1973)}]{Penrose}%
  \BibitemOpen
  \bibfield  {author} {\bibinfo {author} {\bibfnamefont {R.}~\bibnamefont {Penrose}},\ }\href@noop {} {\bibfield  {journal} {\bibinfo  {journal} {Ann. N. Y. Acad. Sci.}\ }\textbf {\bibinfo {volume} {224}},\ \bibinfo {pages} {125} (\bibinfo {year} {1973})}\BibitemShut {NoStop}%
\bibitem [{\citenamefont {Floyd}(1973)}]{Floyd}%
  \BibitemOpen
  \bibfield  {author} {\bibinfo {author} {\bibnamefont {Floyd}},\ }\bibfield  {title} {\enquote {\bibinfo {title} {{The dynamics of Kerr fields}},}\ }\href@noop {} {\bibfield  {journal} {\bibinfo  {journal} {PhD Thesis}\ } (\bibinfo {year} {1973})}\BibitemShut {NoStop}%
\bibitem [{\citenamefont {Carter}(1968)}]{PhysRev.174.1559}%
  \BibitemOpen
  \bibfield  {author} {\bibinfo {author} {\bibfnamefont {Brandon}\ \bibnamefont {Carter}},\ }\bibfield  {title} {\enquote {\bibinfo {title} {Global structure of the kerr family of gravitational fields},}\ }\href {\doibase 10.1103/PhysRev.174.1559} {\bibfield  {journal} {\bibinfo  {journal} {Phys. Rev.}\ }\textbf {\bibinfo {volume} {174}},\ \bibinfo {pages} {1559--1571} (\bibinfo {year} {1968})}\BibitemShut {NoStop}%
\bibitem [{\citenamefont {Frolov}\ \emph {et~al.}(2017)\citenamefont {Frolov}, \citenamefont {Krtous},\ and\ \citenamefont {Kubiznak}}]{Living_Frolov:2017kze}%
  \BibitemOpen
  \bibfield  {author} {\bibinfo {author} {\bibfnamefont {Valeri~P.}\ \bibnamefont {Frolov}}, \bibinfo {author} {\bibfnamefont {Pavel}\ \bibnamefont {Krtous}}, \ and\ \bibinfo {author} {\bibfnamefont {David}\ \bibnamefont {Kubiznak}},\ }\bibfield  {title} {\enquote {\bibinfo {title} {{Black holes, hidden symmetries, and complete integrability}},}\ }\href {\doibase 10.1007/s41114-017-0009-9} {\bibfield  {journal} {\bibinfo  {journal} {Living Rev. Rel.}\ }\textbf {\bibinfo {volume} {20}},\ \bibinfo {pages} {6} (\bibinfo {year} {2017})},\ \Eprint {http://arxiv.org/abs/1705.05482} {arXiv:1705.05482 [gr-qc]} \BibitemShut {NoStop}%
\bibitem [{\citenamefont {{Newman}}\ \emph {et~al.}(1965)\citenamefont {{Newman}}, \citenamefont {{Couch}}, \citenamefont {{Chinnapared}}, \citenamefont {{Exton}}, \citenamefont {{Prakash}},\ and\ \citenamefont {{Torrence}}}]{Kerr-Newman}%
  \BibitemOpen
  \bibfield  {author} {\bibinfo {author} {\bibfnamefont {E.~T.}\ \bibnamefont {{Newman}}}, \bibinfo {author} {\bibfnamefont {E.}~\bibnamefont {{Couch}}}, \bibinfo {author} {\bibfnamefont {K.}~\bibnamefont {{Chinnapared}}}, \bibinfo {author} {\bibfnamefont {A.}~\bibnamefont {{Exton}}}, \bibinfo {author} {\bibfnamefont {A.}~\bibnamefont {{Prakash}}}, \ and\ \bibinfo {author} {\bibfnamefont {R.}~\bibnamefont {{Torrence}}},\ }\bibfield  {title} {\enquote {\bibinfo {title} {{Metric of a Rotating, Charged Mass}},}\ }\href {\doibase 10.1063/1.1704351} {\bibfield  {journal} {\bibinfo  {journal} {Journal of Mathematical Physics}\ }\textbf {\bibinfo {volume} {6}},\ \bibinfo {pages} {918--919} (\bibinfo {year} {1965})}\BibitemShut {NoStop}%
\bibitem [{\citenamefont {Debney}\ \emph {et~al.}(1969)\citenamefont {Debney}, \citenamefont {Kerr},\ and\ \citenamefont {Schild}}]{Kerr-Newman_1}%
  \BibitemOpen
  \bibfield  {author} {\bibinfo {author} {\bibfnamefont {G~C}\ \bibnamefont {Debney}}, \bibinfo {author} {\bibfnamefont {R~P}\ \bibnamefont {Kerr}}, \ and\ \bibinfo {author} {\bibfnamefont {A}~\bibnamefont {Schild}},\ }\bibfield  {title} {\enquote {\bibinfo {title} {Solutions of the einstein and einstein--maxwell equations.}}\ }\href {\doibase 10.1063/1.1664769} {\bibfield  {journal} {\bibinfo  {journal} {J. Math. Phys. (N. Y.), 10: 1842-54(Oct. 1969).}\ }\textbf {\bibinfo {volume} {10}} (\bibinfo {year} {1969}),\ 10.1063/1.1664769}\BibitemShut {NoStop}%
\bibitem [{\citenamefont {{Kerr}}\ and\ \citenamefont {{Schild}}(2009)}]{Kerr_Schild}%
  \BibitemOpen
  \bibfield  {author} {\bibinfo {author} {\bibfnamefont {R.~P.}\ \bibnamefont {{Kerr}}}\ and\ \bibinfo {author} {\bibfnamefont {A.}~\bibnamefont {{Schild}}},\ }\bibfield  {title} {\enquote {\bibinfo {title} {{Republication of: A new class of vacuum solutions of the Einstein field equations}},}\ }\href {\doibase 10.1007/s10714-009-0857-z} {\bibfield  {journal} {\bibinfo  {journal} {General Relativity and Gravitation}\ }\textbf {\bibinfo {volume} {41}},\ \bibinfo {pages} {2485--2499} (\bibinfo {year} {2009})}\BibitemShut {NoStop}%
\bibitem [{\citenamefont {Newman}\ and\ \citenamefont {Janis}(1965)}]{Newman:1965tw}%
  \BibitemOpen
  \bibfield  {author} {\bibinfo {author} {\bibfnamefont {E.~T.}\ \bibnamefont {Newman}}\ and\ \bibinfo {author} {\bibfnamefont {A.~I.}\ \bibnamefont {Janis}},\ }\bibfield  {title} {\enquote {\bibinfo {title} {{Note on the Kerr spinning particle metric}},}\ }\href {\doibase 10.1063/1.1704350} {\bibfield  {journal} {\bibinfo  {journal} {J. Math. Phys.}\ }\textbf {\bibinfo {volume} {6}},\ \bibinfo {pages} {915--917} (\bibinfo {year} {1965})}\BibitemShut {NoStop}%
\bibitem [{\citenamefont {Newman}(1973)}]{Newman:1973afx}%
  \BibitemOpen
  \bibfield  {author} {\bibinfo {author} {\bibfnamefont {E.~T.}\ \bibnamefont {Newman}},\ }\bibfield  {title} {\enquote {\bibinfo {title} {{Complex coordinate transformations and the Schwarzschild-Kerr metrics}},}\ }\href {\doibase 10.1063/1.1666393} {\bibfield  {journal} {\bibinfo  {journal} {J. Math. Phys.}\ }\textbf {\bibinfo {volume} {14}},\ \bibinfo {pages} {774} (\bibinfo {year} {1973})}\BibitemShut {NoStop}%
\bibitem [{\citenamefont {Israel}(1970)}]{Israel}%
  \BibitemOpen
  \bibfield  {author} {\bibinfo {author} {\bibfnamefont {Werner}\ \bibnamefont {Israel}},\ }\bibfield  {title} {\enquote {\bibinfo {title} {Source of the kerr metric},}\ }\href {\doibase 10.1103/PhysRevD.2.641} {\bibfield  {journal} {\bibinfo  {journal} {Phys. Rev. D}\ }\textbf {\bibinfo {volume} {2}},\ \bibinfo {pages} {641--646} (\bibinfo {year} {1970})}\BibitemShut {NoStop}%
\bibitem [{\citenamefont {Kaiser}(2003)}]{Kaiser_2003}%
  \BibitemOpen
  \bibfield  {author} {\bibinfo {author} {\bibfnamefont {Gerald}\ \bibnamefont {Kaiser}},\ }\bibfield  {title} {\enquote {\bibinfo {title} {Physical wavelets and their sources: real physics in complex spacetime},}\ }\href {\doibase 10.1088/0305-4470/36/30/201} {\bibfield  {journal} {\bibinfo  {journal} {Journal of Physics A: Mathematical and General}\ }\textbf {\bibinfo {volume} {36}},\ \bibinfo {pages} {R291--R338} (\bibinfo {year} {2003})}\BibitemShut {NoStop}%
\bibitem [{\citenamefont {Adamo}\ and\ \citenamefont {Newman}(2014)}]{Adamo:2014baa}%
  \BibitemOpen
  \bibfield  {author} {\bibinfo {author} {\bibfnamefont {Tim}\ \bibnamefont {Adamo}}\ and\ \bibinfo {author} {\bibfnamefont {E.~T.}\ \bibnamefont {Newman}},\ }\bibfield  {title} {\enquote {\bibinfo {title} {{The Kerr-Newman metric: A Review}},}\ }\href {\doibase 10.4249/scholarpedia.31791} {\bibfield  {journal} {\bibinfo  {journal} {Scholarpedia}\ }\textbf {\bibinfo {volume} {9}},\ \bibinfo {pages} {31791} (\bibinfo {year} {2014})},\ \Eprint {http://arxiv.org/abs/1410.6626} {arXiv:1410.6626 [gr-qc]} \BibitemShut {NoStop}%
\bibitem [{\citenamefont {Bern}\ \emph {et~al.}(2010)\citenamefont {Bern}, \citenamefont {Carrasco},\ and\ \citenamefont {Johansson}}]{Bern_2010}%
  \BibitemOpen
  \bibfield  {author} {\bibinfo {author} {\bibfnamefont {Zvi}\ \bibnamefont {Bern}}, \bibinfo {author} {\bibfnamefont {John Joseph~M.}\ \bibnamefont {Carrasco}}, \ and\ \bibinfo {author} {\bibfnamefont {Henrik}\ \bibnamefont {Johansson}},\ }\bibfield  {title} {\enquote {\bibinfo {title} {Perturbative quantum gravity as a double copy of gauge theory},}\ }\href {\doibase 10.1103/physrevlett.105.061602} {\bibfield  {journal} {\bibinfo  {journal} {Physical Review Letters}\ }\textbf {\bibinfo {volume} {105}} (\bibinfo {year} {2010}),\ 10.1103/physrevlett.105.061602}\BibitemShut {NoStop}%
\bibitem [{\citenamefont {Monteiro}\ \emph {et~al.}(2014)\citenamefont {Monteiro}, \citenamefont {O'Connell},\ and\ \citenamefont {White}}]{Monteiro_2014}%
  \BibitemOpen
  \bibfield  {author} {\bibinfo {author} {\bibfnamefont {R.}~\bibnamefont {Monteiro}}, \bibinfo {author} {\bibfnamefont {D.}~\bibnamefont {O'Connell}}, \ and\ \bibinfo {author} {\bibfnamefont {C.~D.}\ \bibnamefont {White}},\ }\bibfield  {title} {\enquote {\bibinfo {title} {Black holes and the double copy},}\ }\href {\doibase 10.1007/jhep12(2014)056} {\bibfield  {journal} {\bibinfo  {journal} {Journal of High Energy Physics}\ }\textbf {\bibinfo {volume} {2014}} (\bibinfo {year} {2014}),\ 10.1007/jhep12(2014)056}\BibitemShut {NoStop}%
\bibitem [{\citenamefont {Luna}\ \emph {et~al.}(2015)\citenamefont {Luna}, \citenamefont {Monteiro}, \citenamefont {O'Connell},\ and\ \citenamefont {White}}]{Luna_2015}%
  \BibitemOpen
  \bibfield  {author} {\bibinfo {author} {\bibfnamefont {Andr{\'{e}}s}\ \bibnamefont {Luna}}, \bibinfo {author} {\bibfnamefont {Ricardo}\ \bibnamefont {Monteiro}}, \bibinfo {author} {\bibfnamefont {Donal}\ \bibnamefont {O'Connell}}, \ and\ \bibinfo {author} {\bibfnamefont {Chris~D.}\ \bibnamefont {White}},\ }\bibfield  {title} {\enquote {\bibinfo {title} {The classical double copy for taub{\textendash}{NUT} spacetime},}\ }\href {\doibase 10.1016/j.physletb.2015.09.021} {\bibfield  {journal} {\bibinfo  {journal} {Physics Letters B}\ }\textbf {\bibinfo {volume} {750}},\ \bibinfo {pages} {272--277} (\bibinfo {year} {2015})}\BibitemShut {NoStop}%
\bibitem [{\citenamefont {Bah}\ \emph {et~al.}(2020)\citenamefont {Bah}, \citenamefont {Dempsey},\ and\ \citenamefont {Weck}}]{Bah:2019sda}%
  \BibitemOpen
  \bibfield  {author} {\bibinfo {author} {\bibfnamefont {Ibrahima}\ \bibnamefont {Bah}}, \bibinfo {author} {\bibfnamefont {Ross}\ \bibnamefont {Dempsey}}, \ and\ \bibinfo {author} {\bibfnamefont {Peter}\ \bibnamefont {Weck}},\ }\bibfield  {title} {\enquote {\bibinfo {title} {{Kerr-Schild Double Copy and Complex Worldlines}},}\ }\href {\doibase 10.1007/JHEP02(2020)180} {\bibfield  {journal} {\bibinfo  {journal} {JHEP}\ }\textbf {\bibinfo {volume} {02}},\ \bibinfo {pages} {180} (\bibinfo {year} {2020})},\ \Eprint {http://arxiv.org/abs/1910.04197} {arXiv:1910.04197 [hep-th]} \BibitemShut {NoStop}%
\bibitem [{\citenamefont {White}(2018)}]{White_2018}%
  \BibitemOpen
  \bibfield  {author} {\bibinfo {author} {\bibfnamefont {C.~D.}\ \bibnamefont {White}},\ }\bibfield  {title} {\enquote {\bibinfo {title} {The double copy: gravity from gluons},}\ }\href {\doibase 10.1080/00107514.2017.1415725} {\bibfield  {journal} {\bibinfo  {journal} {Contemporary Physics}\ }\textbf {\bibinfo {volume} {59}},\ \bibinfo {pages} {109--125} (\bibinfo {year} {2018})}\BibitemShut {NoStop}%
\bibitem [{\citenamefont {Bern}\ \emph {et~al.}(2019)\citenamefont {Bern}, \citenamefont {Carrasco}, \citenamefont {Chiodaroli}, \citenamefont {Johansson},\ and\ \citenamefont {Roiban}}]{bern2019duality}%
  \BibitemOpen
  \bibfield  {author} {\bibinfo {author} {\bibfnamefont {Zvi}\ \bibnamefont {Bern}}, \bibinfo {author} {\bibfnamefont {John~Joseph}\ \bibnamefont {Carrasco}}, \bibinfo {author} {\bibfnamefont {Marco}\ \bibnamefont {Chiodaroli}}, \bibinfo {author} {\bibfnamefont {Henrik}\ \bibnamefont {Johansson}}, \ and\ \bibinfo {author} {\bibfnamefont {Radu}\ \bibnamefont {Roiban}},\ }\href@noop {} {\enquote {\bibinfo {title} {The duality between color and kinematics and its applications},}\ } (\bibinfo {year} {2019}),\ \Eprint {http://arxiv.org/abs/1909.01358} {arXiv:1909.01358 [hep-th]} \BibitemShut {NoStop}%
\bibitem [{\citenamefont {Bern}\ \emph {et~al.}(2022)\citenamefont {Bern}, \citenamefont {Carrasco}, \citenamefont {Chiodaroli}, \citenamefont {Johansson},\ and\ \citenamefont {Roiban}}]{bern2022sagex}%
  \BibitemOpen
  \bibfield  {author} {\bibinfo {author} {\bibfnamefont {Zvi}\ \bibnamefont {Bern}}, \bibinfo {author} {\bibfnamefont {John~Joseph}\ \bibnamefont {Carrasco}}, \bibinfo {author} {\bibfnamefont {Marco}\ \bibnamefont {Chiodaroli}}, \bibinfo {author} {\bibfnamefont {Henrik}\ \bibnamefont {Johansson}}, \ and\ \bibinfo {author} {\bibfnamefont {Radu}\ \bibnamefont {Roiban}},\ }\href@noop {} {\enquote {\bibinfo {title} {The sagex review on scattering amplitudes, chapter 2: An invitation to color-kinematics duality and the double copy},}\ } (\bibinfo {year} {2022}),\ \Eprint {http://arxiv.org/abs/2203.13013} {arXiv:2203.13013 [hep-th]} \BibitemShut {NoStop}%
\bibitem [{\citenamefont {de~Paula~Netto}\ \emph {et~al.}(2023)\citenamefont {de~Paula~Netto}, \citenamefont {Giacchini}, \citenamefont {Burzill\`a},\ and\ \citenamefont {Modesto}}]{dePaulaNetto:2023vtg}%
  \BibitemOpen
  \bibfield  {author} {\bibinfo {author} {\bibfnamefont {Tib\'erio}\ \bibnamefont {de~Paula~Netto}}, \bibinfo {author} {\bibfnamefont {Breno~L.}\ \bibnamefont {Giacchini}}, \bibinfo {author} {\bibfnamefont {Nicol\`o}\ \bibnamefont {Burzill\`a}}, \ and\ \bibinfo {author} {\bibfnamefont {Leonardo}\ \bibnamefont {Modesto}},\ }\bibfield  {title} {\enquote {\bibinfo {title} {{Regular black holes from higher-derivative and nonlocal gravity: The smeared delta source approximation}},}\ }\href@noop {} {\  (\bibinfo {year} {2023})},\ \Eprint {http://arxiv.org/abs/2308.12251} {arXiv:2308.12251 [gr-qc]} \BibitemShut {NoStop}%
\bibitem [{\citenamefont {Tomboulis}(1997)}]{Tomboulis:1997gg}%
  \BibitemOpen
  \bibfield  {author} {\bibinfo {author} {\bibfnamefont {E.~T.}\ \bibnamefont {Tomboulis}},\ }\bibfield  {title} {\enquote {\bibinfo {title} {{Superrenormalizable gauge and gravitational theories}},}\ }\href@noop {} {\bibfield  {journal} {\bibinfo  {journal} {{}}\ } (\bibinfo {year} {1997})},\ \Eprint {http://arxiv.org/abs/hep-th/9702146} {arXiv:hep-th/9702146} \BibitemShut {NoStop}%
\bibitem [{\citenamefont {Moffat}(2011)}]{Moffat:2010bh}%
  \BibitemOpen
  \bibfield  {author} {\bibinfo {author} {\bibfnamefont {J.~W.}\ \bibnamefont {Moffat}},\ }\bibfield  {title} {\enquote {\bibinfo {title} {{Ultraviolet Complete Quantum Gravity}},}\ }\href {\doibase 10.1140/epjp/i2011-11043-7} {\bibfield  {journal} {\bibinfo  {journal} {Eur. Phys. J. Plus}\ }\textbf {\bibinfo {volume} {126}},\ \bibinfo {pages} {43} (\bibinfo {year} {2011})},\ \Eprint {http://arxiv.org/abs/1008.2482} {arXiv:1008.2482 [gr-qc]} \BibitemShut {NoStop}%
\bibitem [{\citenamefont {Modesto}(2012)}]{Modesto:2011kw}%
  \BibitemOpen
  \bibfield  {author} {\bibinfo {author} {\bibfnamefont {Leonardo}\ \bibnamefont {Modesto}},\ }\bibfield  {title} {\enquote {\bibinfo {title} {{Super-renormalizable Quantum Gravity}},}\ }\href {\doibase 10.1103/PhysRevD.86.044005} {\bibfield  {journal} {\bibinfo  {journal} {Phys. Rev. D}\ }\textbf {\bibinfo {volume} {86}},\ \bibinfo {pages} {044005} (\bibinfo {year} {2012})},\ \Eprint {http://arxiv.org/abs/1107.2403} {arXiv:1107.2403 [hep-th]} \BibitemShut {NoStop}%
\bibitem [{\citenamefont {Biswas}\ \emph {et~al.}(2012)\citenamefont {Biswas}, \citenamefont {Gerwick}, \citenamefont {Koivisto},\ and\ \citenamefont {Mazumdar}}]{Biswas_2012}%
  \BibitemOpen
  \bibfield  {author} {\bibinfo {author} {\bibfnamefont {Tirthabir}\ \bibnamefont {Biswas}}, \bibinfo {author} {\bibfnamefont {Erik}\ \bibnamefont {Gerwick}}, \bibinfo {author} {\bibfnamefont {Tomi}\ \bibnamefont {Koivisto}}, \ and\ \bibinfo {author} {\bibfnamefont {Anupam}\ \bibnamefont {Mazumdar}},\ }\bibfield  {title} {\enquote {\bibinfo {title} {Towards singularity- and ghost-free theories of gravity},}\ }\href {\doibase 10.1103/physrevlett.108.031101} {\bibfield  {journal} {\bibinfo  {journal} {Physical Review Letters}\ }\textbf {\bibinfo {volume} {108}} (\bibinfo {year} {2012}),\ 10.1103/physrevlett.108.031101}\BibitemShut {NoStop}%
\bibitem [{\citenamefont {Modesto}\ \emph {et~al.}(2011)\citenamefont {Modesto}, \citenamefont {Moffat},\ and\ \citenamefont {Nicolini}}]{Modesto:2010uh}%
  \BibitemOpen
  \bibfield  {author} {\bibinfo {author} {\bibfnamefont {Leonardo}\ \bibnamefont {Modesto}}, \bibinfo {author} {\bibfnamefont {John~W.}\ \bibnamefont {Moffat}}, \ and\ \bibinfo {author} {\bibfnamefont {Piero}\ \bibnamefont {Nicolini}},\ }\bibfield  {title} {\enquote {\bibinfo {title} {{Black holes in an ultraviolet complete quantum gravity}},}\ }\href {\doibase 10.1016/j.physletb.2010.11.046} {\bibfield  {journal} {\bibinfo  {journal} {Phys. Lett. B}\ }\textbf {\bibinfo {volume} {695}},\ \bibinfo {pages} {397--400} (\bibinfo {year} {2011})},\ \Eprint {http://arxiv.org/abs/1010.0680} {arXiv:1010.0680 [gr-qc]} \BibitemShut {NoStop}%
\bibitem [{\citenamefont {Biswas}\ \emph {et~al.}(2013)\citenamefont {Biswas}, \citenamefont {Conroy}, \citenamefont {Koshelev},\ and\ \citenamefont {Mazumdar}}]{Biswas_2013}%
  \BibitemOpen
  \bibfield  {author} {\bibinfo {author} {\bibfnamefont {Tirthabir}\ \bibnamefont {Biswas}}, \bibinfo {author} {\bibfnamefont {Aindri{\'{u} }}\ \bibnamefont {Conroy}}, \bibinfo {author} {\bibfnamefont {Alexey~S}\ \bibnamefont {Koshelev}}, \ and\ \bibinfo {author} {\bibfnamefont {Anupam}\ \bibnamefont {Mazumdar}},\ }\bibfield  {title} {\enquote {\bibinfo {title} {Generalized ghost-free quadratic curvature gravity},}\ }\href {\doibase 10.1088/0264-9381/31/1/015022} {\bibfield  {journal} {\bibinfo  {journal} {Classical and Quantum Gravity}\ }\textbf {\bibinfo {volume} {31}},\ \bibinfo {pages} {015022} (\bibinfo {year} {2013})}\BibitemShut {NoStop}%
\bibitem [{\citenamefont {Boos}\ \emph {et~al.}(2018)\citenamefont {Boos}, \citenamefont {Frolov},\ and\ \citenamefont {Zelnikov}}]{Boos_2018}%
  \BibitemOpen
  \bibfield  {author} {\bibinfo {author} {\bibfnamefont {Jens}\ \bibnamefont {Boos}}, \bibinfo {author} {\bibfnamefont {Valeri~P.}\ \bibnamefont {Frolov}}, \ and\ \bibinfo {author} {\bibfnamefont {Andrei}\ \bibnamefont {Zelnikov}},\ }\bibfield  {title} {\enquote {\bibinfo {title} {Gravitational field of static branes in linearized ghost-free gravity},}\ }\href {\doibase 10.1103/physrevd.97.084021} {\bibfield  {journal} {\bibinfo  {journal} {Physical Review D}\ }\textbf {\bibinfo {volume} {97}} (\bibinfo {year} {2018}),\ 10.1103/physrevd.97.084021}\BibitemShut {NoStop}%
\bibitem [{\citenamefont {Buoninfante}\ \emph {et~al.}(2018)\citenamefont {Buoninfante}, \citenamefont {Cornell}, \citenamefont {Harmsen}, \citenamefont {Koshelev}, \citenamefont {Lambiase}, \citenamefont {Marto},\ and\ \citenamefont {Mazumdar}}]{Buoninfante:2018xif}%
  \BibitemOpen
  \bibfield  {author} {\bibinfo {author} {\bibfnamefont {Luca}\ \bibnamefont {Buoninfante}}, \bibinfo {author} {\bibfnamefont {Alan~S.}\ \bibnamefont {Cornell}}, \bibinfo {author} {\bibfnamefont {Gerhard}\ \bibnamefont {Harmsen}}, \bibinfo {author} {\bibfnamefont {Alexey~S.}\ \bibnamefont {Koshelev}}, \bibinfo {author} {\bibfnamefont {Gaetano}\ \bibnamefont {Lambiase}}, \bibinfo {author} {\bibfnamefont {Jo\~ao}\ \bibnamefont {Marto}}, \ and\ \bibinfo {author} {\bibfnamefont {Anupam}\ \bibnamefont {Mazumdar}},\ }\bibfield  {title} {\enquote {\bibinfo {title} {{Towards nonsingular rotating compact object in ghost-free infinite derivative gravity}},}\ }\href {\doibase 10.1103/PhysRevD.98.084041} {\bibfield  {journal} {\bibinfo  {journal} {Phys. Rev. D}\ }\textbf {\bibinfo {volume} {98}},\ \bibinfo {pages} {084041} (\bibinfo {year} {2018})},\ \Eprint {http://arxiv.org/abs/1807.08896} {arXiv:1807.08896 [gr-qc]} \BibitemShut {NoStop}%
\bibitem [{\citenamefont {Aref'eva}\ \emph {et~al.}(2007)\citenamefont {Aref'eva}, \citenamefont {Joukovskaya},\ and\ \citenamefont {Vernov}}]{Arefeva:2007wvo}%
  \BibitemOpen
  \bibfield  {author} {\bibinfo {author} {\bibfnamefont {I.~Ya.}\ \bibnamefont {Aref'eva}}, \bibinfo {author} {\bibfnamefont {L.~V.}\ \bibnamefont {Joukovskaya}}, \ and\ \bibinfo {author} {\bibfnamefont {S.~Yu.}\ \bibnamefont {Vernov}},\ }\bibfield  {title} {\enquote {\bibinfo {title} {{Bouncing and accelerating solutions in nonlocal stringy models}},}\ }\href {\doibase 10.1088/1126-6708/2007/07/087} {\bibfield  {journal} {\bibinfo  {journal} {JHEP}\ }\textbf {\bibinfo {volume} {07}},\ \bibinfo {pages} {087} (\bibinfo {year} {2007})},\ \Eprint {http://arxiv.org/abs/hep-th/0701184} {arXiv:hep-th/0701184} \BibitemShut {NoStop}%
\bibitem [{\citenamefont {Koshelev}\ and\ \citenamefont {Vernov}(2014)}]{Koshelev:2014voa}%
  \BibitemOpen
  \bibfield  {author} {\bibinfo {author} {\bibfnamefont {Alexey~S.}\ \bibnamefont {Koshelev}}\ and\ \bibinfo {author} {\bibfnamefont {Sergey~Yu.}\ \bibnamefont {Vernov}},\ }\bibfield  {title} {\enquote {\bibinfo {title} {{Cosmological Solutions in Nonlocal Models}},}\ }\href {\doibase 10.1134/S1547477114070255} {\bibfield  {journal} {\bibinfo  {journal} {Phys. Part. Nucl. Lett.}\ }\textbf {\bibinfo {volume} {11}},\ \bibinfo {pages} {960--963} (\bibinfo {year} {2014})},\ \Eprint {http://arxiv.org/abs/1406.5887} {arXiv:1406.5887 [gr-qc]} \BibitemShut {NoStop}%
\bibitem [{\citenamefont {Kilicarslan}(2019)}]{Kilicarslan:2019njc}%
  \BibitemOpen
  \bibfield  {author} {\bibinfo {author} {\bibfnamefont {Ercan}\ \bibnamefont {Kilicarslan}},\ }\bibfield  {title} {\enquote {\bibinfo {title} {{$pp$-waves as exact solutions to ghost-free infinite derivative gravity}},}\ }\href {\doibase 10.1103/PhysRevD.99.124048} {\bibfield  {journal} {\bibinfo  {journal} {Phys. Rev. D}\ }\textbf {\bibinfo {volume} {99}},\ \bibinfo {pages} {124048} (\bibinfo {year} {2019})},\ \Eprint {http://arxiv.org/abs/1903.04283} {arXiv:1903.04283 [gr-qc]} \BibitemShut {NoStop}%
\bibitem [{\citenamefont {Dengiz}\ \emph {et~al.}(2020)\citenamefont {Dengiz}, \citenamefont {Kilicarslan}, \citenamefont {Kol\'a\v{r}},\ and\ \citenamefont {Mazumdar}}]{Dengiz:2020xbu}%
  \BibitemOpen
  \bibfield  {author} {\bibinfo {author} {\bibfnamefont {Suat}\ \bibnamefont {Dengiz}}, \bibinfo {author} {\bibfnamefont {Ercan}\ \bibnamefont {Kilicarslan}}, \bibinfo {author} {\bibfnamefont {Ivan}\ \bibnamefont {Kol\'a\v{r}}}, \ and\ \bibinfo {author} {\bibfnamefont {Anupam}\ \bibnamefont {Mazumdar}},\ }\bibfield  {title} {\enquote {\bibinfo {title} {{Impulsive waves in ghost free infinite derivative gravity in anti-de Sitter spacetime}},}\ }\href {\doibase 10.1103/PhysRevD.102.044016} {\bibfield  {journal} {\bibinfo  {journal} {Phys. Rev. D}\ }\textbf {\bibinfo {volume} {102}},\ \bibinfo {pages} {044016} (\bibinfo {year} {2020})},\ \Eprint {http://arxiv.org/abs/2006.07650} {arXiv:2006.07650 [gr-qc]} \BibitemShut {NoStop}%
\bibitem [{\citenamefont {Kol\'a\v{r}}\ \emph {et~al.}(2022)\citenamefont {Kol\'a\v{r}}, \citenamefont {M\'alek}, \citenamefont {Dengiz},\ and\ \citenamefont {Kilicarslan}}]{Kolar:2021uiu}%
  \BibitemOpen
  \bibfield  {author} {\bibinfo {author} {\bibfnamefont {Ivan}\ \bibnamefont {Kol\'a\v{r}}}, \bibinfo {author} {\bibfnamefont {Tom\'a\v{s}}\ \bibnamefont {M\'alek}}, \bibinfo {author} {\bibfnamefont {Suat}\ \bibnamefont {Dengiz}}, \ and\ \bibinfo {author} {\bibfnamefont {Ercan}\ \bibnamefont {Kilicarslan}},\ }\bibfield  {title} {\enquote {\bibinfo {title} {{Exact gyratons in higher and infinite derivative gravity}},}\ }\href {\doibase 10.1103/PhysRevD.105.044018} {\bibfield  {journal} {\bibinfo  {journal} {Phys. Rev. D}\ }\textbf {\bibinfo {volume} {105}},\ \bibinfo {pages} {044018} (\bibinfo {year} {2022})},\ \Eprint {http://arxiv.org/abs/2107.11884} {arXiv:2107.11884 [gr-qc]} \BibitemShut {NoStop}%
\bibitem [{\citenamefont {Boos}(2020)}]{Boos:2020qgg}%
  \BibitemOpen
  \bibfield  {author} {\bibinfo {author} {\bibfnamefont {Jens}\ \bibnamefont {Boos}},\ }\emph {\bibinfo {title} {{Effects of Non-locality in Gravity and Quantum Theory}}},\ \href {\doibase 10.7939/r3-7bt0-na76} {Ph.D. thesis},\ \bibinfo  {school} {Alberta U.} (\bibinfo {year} {2020}),\ \Eprint {http://arxiv.org/abs/2009.10856} {arXiv:2009.10856 [gr-qc]} \BibitemShut {NoStop}%
\bibitem [{\citenamefont {Modesto}\ and\ \citenamefont {Rachwa\l{}}(2017)}]{Modesto:2017sdr}%
  \BibitemOpen
  \bibfield  {author} {\bibinfo {author} {\bibfnamefont {Leonardo}\ \bibnamefont {Modesto}}\ and\ \bibinfo {author} {\bibfnamefont {Les\l{}aw}\ \bibnamefont {Rachwa\l{}}},\ }\bibfield  {title} {\enquote {\bibinfo {title} {{Nonlocal quantum gravity: A review}},}\ }\href {\doibase 10.1142/S0218271817300208} {\bibfield  {journal} {\bibinfo  {journal} {Int. J. Mod. Phys. D}\ }\textbf {\bibinfo {volume} {26}},\ \bibinfo {pages} {1730020} (\bibinfo {year} {2017})}\BibitemShut {NoStop}%
\bibitem [{\citenamefont {Buoninfante}(2019)}]{Buoninfante:2019zws}%
  \BibitemOpen
  \bibfield  {author} {\bibinfo {author} {\bibfnamefont {Luca}\ \bibnamefont {Buoninfante}},\ }\emph {\bibinfo {title} {Nonlocal Field theories: Theoretical and Phenomenological Aspects}},\ \href {\doibase 10.33612/diss.99349099} {Ph.D. thesis},\ \bibinfo  {school} {University of Groningen} (\bibinfo {year} {2019})\BibitemShut {NoStop}%
\bibitem [{\citenamefont {Heredia}\ \emph {et~al.}(2022)\citenamefont {Heredia}, \citenamefont {Kol\'a\v{r}}, \citenamefont {Llosa}, \citenamefont {Torralba},\ and\ \citenamefont {Mazumdar}}]{Heredia:2021pxp}%
  \BibitemOpen
  \bibfield  {author} {\bibinfo {author} {\bibfnamefont {Carlos}\ \bibnamefont {Heredia}}, \bibinfo {author} {\bibfnamefont {Ivan}\ \bibnamefont {Kol\'a\v{r}}}, \bibinfo {author} {\bibfnamefont {Josep}\ \bibnamefont {Llosa}}, \bibinfo {author} {\bibfnamefont {Francisco Jos\'e~Maldonado}\ \bibnamefont {Torralba}}, \ and\ \bibinfo {author} {\bibfnamefont {Anupam}\ \bibnamefont {Mazumdar}},\ }\bibfield  {title} {\enquote {\bibinfo {title} {{Infinite-derivative linearized gravity in convolutional form}},}\ }\href {\doibase 10.1088/1361-6382/ac5a14} {\bibfield  {journal} {\bibinfo  {journal} {Class. Quant. Grav.}\ }\textbf {\bibinfo {volume} {39}},\ \bibinfo {pages} {085001} (\bibinfo {year} {2022})},\ \Eprint {http://arxiv.org/abs/2112.05397} {arXiv:2112.05397 [gr-qc]} \BibitemShut {NoStop}%
\bibitem [{\citenamefont {Kol\'a\v{r}}(2022)}]{Kolar:2022kgx}%
  \BibitemOpen
  \bibfield  {author} {\bibinfo {author} {\bibfnamefont {Ivan}\ \bibnamefont {Kol\'a\v{r}}},\ }\bibfield  {title} {\enquote {\bibinfo {title} {{Nonlocal scalar fields in static spacetimes via heat kernels}},}\ }\href {\doibase 10.1103/PhysRevD.105.084026} {\bibfield  {journal} {\bibinfo  {journal} {Phys. Rev. D}\ }\textbf {\bibinfo {volume} {105}},\ \bibinfo {pages} {084026} (\bibinfo {year} {2022})},\ \Eprint {http://arxiv.org/abs/2201.09908} {arXiv:2201.09908 [gr-qc]} \BibitemShut {NoStop}%
\bibitem [{\citenamefont {Buoninfante}\ \emph {et~al.}(2022)\citenamefont {Buoninfante}, \citenamefont {Giacchini},\ and\ \citenamefont {de~Paula~Netto}}]{Buoninfante:2022ild}%
  \BibitemOpen
  \bibfield  {author} {\bibinfo {author} {\bibfnamefont {Luca}\ \bibnamefont {Buoninfante}}, \bibinfo {author} {\bibfnamefont {Breno~L.}\ \bibnamefont {Giacchini}}, \ and\ \bibinfo {author} {\bibfnamefont {Tib\'erio}\ \bibnamefont {de~Paula~Netto}},\ }\bibfield  {title} {\enquote {\bibinfo {title} {{Black holes in non-local gravity}},}\ }\href@noop {} {\bibfield  {journal} {\bibinfo  {journal} {{}}\ } (\bibinfo {year} {2022})},\ \Eprint {http://arxiv.org/abs/2211.03497} {arXiv:2211.03497 [gr-qc]} \BibitemShut {NoStop}%
\bibitem [{\citenamefont {Kol\'a\v{r}}\ \emph {et~al.}(2021)\citenamefont {Kol\'a\v{r}}, \citenamefont {M\'alek},\ and\ \citenamefont {Mazumdar}}]{Kolar:2021rfl}%
  \BibitemOpen
  \bibfield  {author} {\bibinfo {author} {\bibfnamefont {Ivan}\ \bibnamefont {Kol\'a\v{r}}}, \bibinfo {author} {\bibfnamefont {Tom\'a\v{s}}\ \bibnamefont {M\'alek}}, \ and\ \bibinfo {author} {\bibfnamefont {Anupam}\ \bibnamefont {Mazumdar}},\ }\bibfield  {title} {\enquote {\bibinfo {title} {{Exact solutions of nonlocal gravity in a class of almost universal spacetimes}},}\ }\href {\doibase 10.1103/PhysRevD.103.124067} {\bibfield  {journal} {\bibinfo  {journal} {Phys. Rev. D}\ }\textbf {\bibinfo {volume} {103}},\ \bibinfo {pages} {124067} (\bibinfo {year} {2021})},\ \Eprint {http://arxiv.org/abs/2103.08555} {arXiv:2103.08555 [gr-qc]} \BibitemShut {NoStop}%
\bibitem [{\citenamefont {Visser}(2007)}]{Visser:2007fj}%
  \BibitemOpen
  \bibfield  {author} {\bibinfo {author} {\bibfnamefont {Matt}\ \bibnamefont {Visser}},\ }\bibfield  {title} {\enquote {\bibinfo {title} {{The Kerr spacetime: A Brief introduction}},}\ }in\ \href@noop {} {\emph {\bibinfo {booktitle} {{Kerr Fest: Black Holes in Astrophysics, General Relativity and Quantum Gravity}}}}\ (\bibinfo {year} {2007})\ \Eprint {http://arxiv.org/abs/0706.0622} {arXiv:0706.0622 [gr-qc]} \BibitemShut {NoStop}%
\bibitem [{\citenamefont {Geroch}(1972)}]{Geroch_2}%
  \BibitemOpen
  \bibfield  {author} {\bibinfo {author} {\bibfnamefont {R.~P.}\ \bibnamefont {Geroch}},\ }\bibfield  {title} {\enquote {\bibinfo {title} {{A Method for generating new solutions of Einstein's equation. 2}},}\ }\href {\doibase 10.1063/1.1665990} {\bibfield  {journal} {\bibinfo  {journal} {J. Math. Phys.}\ }\textbf {\bibinfo {volume} {13}},\ \bibinfo {pages} {394--404} (\bibinfo {year} {1972})}\BibitemShut {NoStop}%
\bibitem [{\citenamefont {Frolov}\ and\ \citenamefont {Zelnikov}(2011)}]{FrolovZelnikov:2011}%
  \BibitemOpen
  \bibfield  {author} {\bibinfo {author} {\bibfnamefont {Valeri~P.}\ \bibnamefont {Frolov}}\ and\ \bibinfo {author} {\bibfnamefont {A.}~\bibnamefont {Zelnikov}},\ }\href {http://books.google.ca/books?id=UTGFNAEACAAJ} {\emph {\bibinfo {title} {{Introduction to Black Hole Physics}}}}\ (\bibinfo  {publisher} {Oxford University Press},\ \bibinfo {year} {2011})\BibitemShut {NoStop}%
\bibitem [{\citenamefont {Sommers}(1976)}]{sommers1976}%
  \BibitemOpen
  \bibfield  {author} {\bibinfo {author} {\bibfnamefont {P}~\bibnamefont {Sommers}},\ }\bibfield  {title} {\enquote {\bibinfo {title} {Properties of shear-free congruences of null geodesics},}\ }\href@noop {} {\bibfield  {journal} {\bibinfo  {journal} {Proceedings of the Royal Society of London. A. Mathematical and Physical Sciences}\ }\textbf {\bibinfo {volume} {349}},\ \bibinfo {pages} {309--318} (\bibinfo {year} {1976})}\BibitemShut {NoStop}%
\bibitem [{\citenamefont {Frolov}(1979)}]{Frolov1979}%
  \BibitemOpen
  \bibfield  {author} {\bibinfo {author} {\bibfnamefont {V.~P.}\ \bibnamefont {Frolov}},\ }\enquote {\bibinfo {title} {The newman-penrose method in the theory of general relativity},}\ in\ \href {\doibase 10.1007/978-1-4684-0676-4_4} {\emph {\bibinfo {booktitle} {Problems in the General Theory of Relativity and Theory of Group Representations}}},\ \bibinfo {editor} {edited by\ \bibinfo {editor} {\bibfnamefont {N.~G.}\ \bibnamefont {Basov}}}\ (\bibinfo  {publisher} {Springer US},\ \bibinfo {address} {Boston, MA},\ \bibinfo {year} {1979})\ pp.\ \bibinfo {pages} {73--185}\BibitemShut {NoStop}%
\bibitem [{\citenamefont {Newman}(2004)}]{Newman_2004}%
  \BibitemOpen
  \bibfield  {author} {\bibinfo {author} {\bibfnamefont {Ezra~T}\ \bibnamefont {Newman}},\ }\bibfield  {title} {\enquote {\bibinfo {title} {Maxwell fields and shear-free null geodesic congruences},}\ }\href {\doibase 10.1088/0264-9381/21/13/007} {\bibfield  {journal} {\bibinfo  {journal} {Classical and Quantum Gravity}\ }\textbf {\bibinfo {volume} {21}},\ \bibinfo {pages} {3197--3221} (\bibinfo {year} {2004})}\BibitemShut {NoStop}%
\bibitem [{\citenamefont {Brewster}\ and\ \citenamefont {Franson}(2018)}]{Brewster_2018}%
  \BibitemOpen
  \bibfield  {author} {\bibinfo {author} {\bibfnamefont {R.~A.}\ \bibnamefont {Brewster}}\ and\ \bibinfo {author} {\bibfnamefont {J.~D.}\ \bibnamefont {Franson}},\ }\bibfield  {title} {\enquote {\bibinfo {title} {Generalized delta functions and their use in quantum optics},}\ }\href {\doibase 10.1063/1.4985938} {\bibfield  {journal} {\bibinfo  {journal} {Journal of Mathematical Physics}\ }\textbf {\bibinfo {volume} {59}},\ \bibinfo {pages} {012102} (\bibinfo {year} {2018})}\BibitemShut {NoStop}%
\bibitem [{\citenamefont {Lindell}(1993)}]{LINDELL}%
  \BibitemOpen
  \bibfield  {author} {\bibinfo {author} {\bibfnamefont {Ismo}\ \bibnamefont {Lindell}},\ }\bibfield  {title} {\enquote {\bibinfo {title} {Delta function expansions, complex delta functions and the steepest descent method},}\ }\href {\doibase 10.1119/1.17238} {\bibfield  {journal} {\bibinfo  {journal} {American Journal of Physics - AMER J PHYS}\ }\textbf {\bibinfo {volume} {61}},\ \bibinfo {pages} {438--442} (\bibinfo {year} {1993})}\BibitemShut {NoStop}%
\bibitem [{\citenamefont {Smagin}(2014)}]{Smagin}%
  \BibitemOpen
  \bibfield  {author} {\bibinfo {author} {\bibfnamefont {V.~A.}\ \bibnamefont {Smagin}},\ }\bibfield  {title} {\enquote {\bibinfo {title} {Complex delta function and its information application},}\ }\href {\doibase 10.3103/S0146411614010064} {\bibfield  {journal} {\bibinfo  {journal} {Autom. Control. Comput. Sci.}\ }\textbf {\bibinfo {volume} {48}},\ \bibinfo {pages} {10--16} (\bibinfo {year} {2014})}\BibitemShut {NoStop}%
\bibitem [{\citenamefont {Berry}(2017)}]{Berry}%
  \BibitemOpen
  \bibfield  {author} {\bibinfo {author} {\bibfnamefont {Michael}\ \bibnamefont {Berry}},\ }\enquote {\bibinfo {title} {Faster than fourier},}\ in\ \href {\doibase 10.1142/10480} {\emph {\bibinfo {booktitle} {A Half-century of Physical Asymptotics and Other Diversions}}}\ (\bibinfo  {publisher} {WORLD SCIENTIFIC},\ \bibinfo {year} {2017})\ pp.\ \bibinfo {pages} {483--493}\BibitemShut {NoStop}%
\bibitem [{\citenamefont {Oldham}\ \emph {et~al.}(2010)\citenamefont {Oldham}, \citenamefont {Myland},\ and\ \citenamefont {Spanier}}]{oldham2010atlas}%
  \BibitemOpen
  \bibfield  {author} {\bibinfo {author} {\bibfnamefont {K.B.}\ \bibnamefont {Oldham}}, \bibinfo {author} {\bibfnamefont {J.}~\bibnamefont {Myland}}, \ and\ \bibinfo {author} {\bibfnamefont {J.}~\bibnamefont {Spanier}},\ }\href {https://books.google.ca/books?id=UrSnNeJW10YC} {\emph {\bibinfo {title} {An Atlas of Functions: with Equator, the Atlas Function Calculator}}},\ An Atlas of Functions\ (\bibinfo  {publisher} {Springer New York},\ \bibinfo {year} {2010})\BibitemShut {NoStop}%
\bibitem [{\citenamefont {Kaiser}(2004)}]{Kaiser_2004}%
  \BibitemOpen
  \bibfield  {author} {\bibinfo {author} {\bibfnamefont {Gerald}\ \bibnamefont {Kaiser}},\ }\bibfield  {title} {\enquote {\bibinfo {title} {Distributional sources for newman's holomorphic coulomb field},}\ }\href {\doibase 10.1088/0305-4470/37/36/011} {\bibfield  {journal} {\bibinfo  {journal} {Journal of Physics A: Mathematical and General}\ }\textbf {\bibinfo {volume} {37}},\ \bibinfo {pages} {8735--8745} (\bibinfo {year} {2004})}\BibitemShut {NoStop}%
\bibitem [{\citenamefont {Eleni}\ and\ \citenamefont {Apostolatos}(2020)}]{Eleni_2020}%
  \BibitemOpen
  \bibfield  {author} {\bibinfo {author} {\bibfnamefont {Areti}\ \bibnamefont {Eleni}}\ and\ \bibinfo {author} {\bibfnamefont {Theocharis~A.}\ \bibnamefont {Apostolatos}},\ }\bibfield  {title} {\enquote {\bibinfo {title} {Newtonian analogue of a kerr black hole},}\ }\href {\doibase 10.1103/physrevd.101.044056} {\bibfield  {journal} {\bibinfo  {journal} {Physical Review D}\ }\textbf {\bibinfo {volume} {101}} (\bibinfo {year} {2020}),\ 10.1103/physrevd.101.044056}\BibitemShut {NoStop}%
\bibitem [{\citenamefont {Frolov}(2015)}]{Frolov:2015bta}%
  \BibitemOpen
  \bibfield  {author} {\bibinfo {author} {\bibfnamefont {Valeri~P.}\ \bibnamefont {Frolov}},\ }\bibfield  {title} {\enquote {\bibinfo {title} {{Mass-gap for black hole formation in higher derivative and ghost free gravity}},}\ }\href {\doibase 10.1103/PhysRevLett.115.051102} {\bibfield  {journal} {\bibinfo  {journal} {Phys. Rev. Lett.}\ }\textbf {\bibinfo {volume} {115}},\ \bibinfo {pages} {051102} (\bibinfo {year} {2015})},\ \Eprint {http://arxiv.org/abs/1505.00492} {arXiv:1505.00492 [hep-th]} \BibitemShut {NoStop}%
\bibitem [{\citenamefont {Boos}\ \emph {et~al.}(2020)\citenamefont {Boos}, \citenamefont {Pinedo~Soto},\ and\ \citenamefont {Frolov}}]{Boos:2020}%
  \BibitemOpen
  \bibfield  {author} {\bibinfo {author} {\bibfnamefont {Jens}\ \bibnamefont {Boos}}, \bibinfo {author} {\bibfnamefont {Jose}\ \bibnamefont {Pinedo~Soto}}, \ and\ \bibinfo {author} {\bibfnamefont {Valeri~P.}\ \bibnamefont {Frolov}},\ }\bibfield  {title} {\enquote {\bibinfo {title} {{Ultrarelativistic spinning objects in nonlocal ghost-free gravity}},}\ }\href {\doibase 10.1103/PhysRevD.101.124065} {\bibfield  {journal} {\bibinfo  {journal} {Phys. Rev. D}\ }\textbf {\bibinfo {volume} {101}},\ \bibinfo {pages} {124065} (\bibinfo {year} {2020})},\ \Eprint {http://arxiv.org/abs/2004.07420} {arXiv:2004.07420 [gr-qc]} \BibitemShut {NoStop}%
\bibitem [{\citenamefont {DeWitt}(1965)}]{dewitt1965dynamical}%
  \BibitemOpen
  \bibfield  {author} {\bibinfo {author} {\bibfnamefont {B.S.}\ \bibnamefont {DeWitt}},\ }\href {https://books.google.ca/books?id=0vQyAAAAMAAJ} {\emph {\bibinfo {title} {Dynamical Theory of Groups and Fields}}},\ Documents on modern physics\ (\bibinfo  {publisher} {Gordon and Breach},\ \bibinfo {year} {1965})\BibitemShut {NoStop}%
\bibitem [{\citenamefont {DeWitt}(1975)}]{DeWitt:1975ys}%
  \BibitemOpen
  \bibfield  {author} {\bibinfo {author} {\bibfnamefont {Bryce~S.}\ \bibnamefont {DeWitt}},\ }\bibfield  {title} {\enquote {\bibinfo {title} {{Quantum Field Theory in Curved Space-Time}},}\ }\href {\doibase 10.1016/0370-1573(75)90051-4} {\bibfield  {journal} {\bibinfo  {journal} {Phys. Rept.}\ }\textbf {\bibinfo {volume} {19}},\ \bibinfo {pages} {295--357} (\bibinfo {year} {1975})}\BibitemShut {NoStop}%
\bibitem [{\citenamefont {Abramowitz}\ and\ \citenamefont {Stegun}(1965)}]{abramowitz}%
  \BibitemOpen
  \bibfield  {author} {\bibinfo {author} {\bibfnamefont {M.}~\bibnamefont {Abramowitz}}\ and\ \bibinfo {author} {\bibfnamefont {I.A.}\ \bibnamefont {Stegun}},\ }\href {https://books.google.ca/books?id=MtU8uP7XMvoC} {\emph {\bibinfo {title} {Handbook of Mathematical Functions: With Formulas, Graphs, and Mathematical Tables}}},\ Applied mathematics series\ (\bibinfo  {publisher} {Dover Publications},\ \bibinfo {year} {1965})\BibitemShut {NoStop}%
\bibitem [{\citenamefont {Senovilla}(2011)}]{Senovilla:2011fk}%
  \BibitemOpen
  \bibfield  {author} {\bibinfo {author} {\bibfnamefont {Jose M.~M.}\ \bibnamefont {Senovilla}},\ }\bibfield  {title} {\enquote {\bibinfo {title} {{Trapped surfaces}},}\ }\href {\doibase 10.1142/S0218271811020354} {\bibfield  {journal} {\bibinfo  {journal} {Int. J. Mod. Phys. D}\ }\textbf {\bibinfo {volume} {20}},\ \bibinfo {pages} {2139} (\bibinfo {year} {2011})},\ \Eprint {http://arxiv.org/abs/1107.1344} {arXiv:1107.1344 [gr-qc]} \BibitemShut {NoStop}%
\bibitem [{\citenamefont {Flammer}(2014)}]{flammer}%
  \BibitemOpen
  \bibfield  {author} {\bibinfo {author} {\bibfnamefont {Carson}\ \bibnamefont {Flammer}},\ }\href@noop {} {\emph {\bibinfo {title} {Spheroidal wave functions}}}\ (\bibinfo  {publisher} {Courier Corporation},\ \bibinfo {year} {2014})\BibitemShut {NoStop}%
\bibitem [{\citenamefont {Olver}\ \emph {et~al.}(2010)\citenamefont {Olver}, \citenamefont {Lozier}, \citenamefont {Boisvert},\ and\ \citenamefont {Clark}}]{NIST}%
  \BibitemOpen
  \bibfield  {author} {\bibinfo {author} {\bibfnamefont {Frank}\ \bibnamefont {Olver}}, \bibinfo {author} {\bibfnamefont {Daniel}\ \bibnamefont {Lozier}}, \bibinfo {author} {\bibfnamefont {Ronald}\ \bibnamefont {Boisvert}}, \ and\ \bibinfo {author} {\bibfnamefont {Charles}\ \bibnamefont {Clark}},\ }\href@noop {} {\emph {\bibinfo {title} {The NIST Handbook of Mathematical Functions}}}\ (\bibinfo  {publisher} {Cambridge University Press, New York, NY},\ \bibinfo {year} {2010})\BibitemShut {NoStop}%
\bibitem [{\citenamefont {Lee}\ \emph {et~al.}(2020)\citenamefont {Lee}, \citenamefont {Adelberger}, \citenamefont {Cook}, \citenamefont {Fleischer},\ and\ \citenamefont {Heckel}}]{Lee:2020zjt}%
  \BibitemOpen
  \bibfield  {author} {\bibinfo {author} {\bibfnamefont {J.~G.}\ \bibnamefont {Lee}}, \bibinfo {author} {\bibfnamefont {E.~G.}\ \bibnamefont {Adelberger}}, \bibinfo {author} {\bibfnamefont {T.~S.}\ \bibnamefont {Cook}}, \bibinfo {author} {\bibfnamefont {S.~M.}\ \bibnamefont {Fleischer}}, \ and\ \bibinfo {author} {\bibfnamefont {B.~R.}\ \bibnamefont {Heckel}},\ }\bibfield  {title} {\enquote {\bibinfo {title} {{New Test of the Gravitational $1/r^2$ Law at Separations down to 52 $\mu$m}},}\ }\href {\doibase 10.1103/PhysRevLett.124.101101} {\bibfield  {journal} {\bibinfo  {journal} {Phys. Rev. Lett.}\ }\textbf {\bibinfo {volume} {124}},\ \bibinfo {pages} {101101} (\bibinfo {year} {2020})},\ \Eprint {http://arxiv.org/abs/2002.11761} {arXiv:2002.11761 [hep-ex]} \BibitemShut {NoStop}%
\bibitem [{\citenamefont {Gurses}\ and\ \citenamefont {Feza}(1975)}]{Gurses:1975vu}%
  \BibitemOpen
  \bibfield  {author} {\bibinfo {author} {\bibfnamefont {Metin}\ \bibnamefont {Gurses}}\ and\ \bibinfo {author} {\bibfnamefont {Gursey}\ \bibnamefont {Feza}},\ }\bibfield  {title} {\enquote {\bibinfo {title} {{Lorentz Covariant Treatment of the Kerr-Schild Metric}},}\ }\href {\doibase 10.1063/1.522480} {\bibfield  {journal} {\bibinfo  {journal} {J. Math. Phys.}\ }\textbf {\bibinfo {volume} {16}},\ \bibinfo {pages} {2385} (\bibinfo {year} {1975})}\BibitemShut {NoStop}%
\bibitem [{\citenamefont {Babichev}\ \emph {et~al.}(2020)\citenamefont {Babichev}, \citenamefont {Charmousis}, \citenamefont {Cisterna},\ and\ \citenamefont {Hassaine}}]{Babichev:2020qpr}%
  \BibitemOpen
  \bibfield  {author} {\bibinfo {author} {\bibfnamefont {Eugeny}\ \bibnamefont {Babichev}}, \bibinfo {author} {\bibfnamefont {Christos}\ \bibnamefont {Charmousis}}, \bibinfo {author} {\bibfnamefont {Adolfo}\ \bibnamefont {Cisterna}}, \ and\ \bibinfo {author} {\bibfnamefont {Mokhtar}\ \bibnamefont {Hassaine}},\ }\bibfield  {title} {\enquote {\bibinfo {title} {{Regular black holes via the Kerr-Schild construction in DHOST theories}},}\ }\href {\doibase 10.1088/1475-7516/2020/06/049} {\bibfield  {journal} {\bibinfo  {journal} {JCAP}\ }\textbf {\bibinfo {volume} {06}},\ \bibinfo {pages} {049} (\bibinfo {year} {2020})},\ \Eprint {http://arxiv.org/abs/2004.00597} {arXiv:2004.00597 [hep-th]} \BibitemShut {NoStop}%
\bibitem [{\citenamefont {Torres}(2022)}]{torres2022regular}%
  \BibitemOpen
  \bibfield  {author} {\bibinfo {author} {\bibfnamefont {Ramón}\ \bibnamefont {Torres}},\ }\href@noop {} {\enquote {\bibinfo {title} {Regular rotating black holes: A review},}\ } (\bibinfo {year} {2022}),\ \Eprint {http://arxiv.org/abs/2208.12713} {arXiv:2208.12713 [gr-qc]} \BibitemShut {NoStop}%
\bibitem [{\citenamefont {Baines}\ and\ \citenamefont {Visser}(2023)}]{Baines:2023dhq}%
  \BibitemOpen
  \bibfield  {author} {\bibinfo {author} {\bibfnamefont {Joshua}\ \bibnamefont {Baines}}\ and\ \bibinfo {author} {\bibfnamefont {Matt}\ \bibnamefont {Visser}},\ }\bibfield  {title} {\enquote {\bibinfo {title} {{Killing Horizons and Surface Gravities for a Well-Behaved Three-Function Generalization of the Kerr Spacetime}},}\ }\href {\doibase 10.3390/universe9050223} {\bibfield  {journal} {\bibinfo  {journal} {Universe}\ }\textbf {\bibinfo {volume} {9}},\ \bibinfo {pages} {223} (\bibinfo {year} {2023})},\ \Eprint {http://arxiv.org/abs/2303.07380} {arXiv:2303.07380 [gr-qc]} \BibitemShut {NoStop}%
\bibitem [{\citenamefont {Zhou}\ and\ \citenamefont {Modesto}(2023)}]{Zhou:2023lwc}%
  \BibitemOpen
  \bibfield  {author} {\bibinfo {author} {\bibfnamefont {Tian}\ \bibnamefont {Zhou}}\ and\ \bibinfo {author} {\bibfnamefont {Leonardo}\ \bibnamefont {Modesto}},\ }\bibfield  {title} {\enquote {\bibinfo {title} {{On the analytic extension of regular rotating black holes}},}\ }\href@noop {} {\bibfield  {journal} {\bibinfo  {journal} {{}}\ } (\bibinfo {year} {2023})},\ \Eprint {http://arxiv.org/abs/2303.11322} {arXiv:2303.11322 [gr-qc]} \BibitemShut {NoStop}%
\end{thebibliography}%

\end{document}